\documentclass[12pt,a4paper]{iopjournal}

\usepackage{graphicx}
\usepackage{caption}
\usepackage{subcaption}
\usepackage{siunitx}
\usepackage{verbatim}
\DeclareSIUnit{\au}{a.u.}
\usepackage{bm}% bold math
\usepackage{float}

\newcommand{\sd}{SiO\textsubscript{2}}

\newcommand{\etal}{{\it et al\/}\ }

\begin{document}
\articletype {Paper}
\title{Multi-diagnostic characterization of inductively coupled discharges with tailored waveform substrate bias for precise control of plasma etching} 

\author{Jonas Giesekus$^{1}$\orcid{0000-0002-4506-867X
}, Anton Pletzer$^{2}$\orcid{0009-0003-0414-8643}, Florian Beckfeld$^{1}$\orcid{0000-0001-8605-2634}, Katharina Noesges$^{1}$\orcid{0000-0002-1579-3301}, Claudia Bock$^{2}$\orcid{0000-0003-0933-9556},  Julian~Schulze$^{1}$\orcid{0000-0001-7929-5734}}

\affil{$^{1}$ Chair of Applied Electrodynamics and Plasma Technology, Faculty of Electrical Engineering and Information Technology, Ruhr-University Bochum, D-44780 Bochum, Germany}

\affil{$^{2}$ Chair of Microsystems Technology, Faculty of Electrical Engineering and Information Technology, Ruhr-University Bochum, D-44780 Bochum, Germany }

\email{giesekus@aept.rub.de}
\begin{abstract}
Precise control of ion energy distribution functions (IEDF) is crucial for selectivity as well as control over sputter rate and substrate damage in nanoscale plasma processes. In this work, a low frequency (\SI{100}{\kilo\hertz}) tailored pulse-wave-shaped bias voltage waveform is applied to the substrate electrode of an inductively coupled plasma (ICP) and its effects on the IEDF, electron density, electron dynamics and the etch rates of silicon dioxide as well as amorphous silicon are investigated in a commercial \SI{200}{\milli\meter} reactive ion etching (RIE) reactor. While the tailored waveform substrate bias hardly affects the electron density above the substrate and the spatio-temporally resolved electron power absorption dynamics, it is found to affect the ion flux to the substrate at high ICP source powers. Monoenergetic IEDFs with a full width at half maximum (FWHM) below \SI{10}{\electronvolt} are realized with mean ion energies ranging from \SI{20}{\electronvolt} to \SI{100}{\electronvolt} in both argon and SF\textsubscript{6}. Such monoenergetic IEDFs are used to determine the Ar ion sputter threshold energies of amorphous silicon and silicon dioxide to be \SI{23}{\electronvolt} and \SI{37}{\electronvolt}, respectively, and to realize selective etching of these two materials by Ar ion sputtering based on tailoring the IEDF to ensure that all incident ions are within this narrow ion energy selectivity window.

\noindent
\end{abstract}

\\

\section{Introduction}

After 50 years of Moore's law driving exponential growth in semiconductor performance, the industry is now approaching a fundamental limit. The critical features of semiconductors have been reduced to just a few atoms in thickness, necessitating new design approaches to further enhance computing power \cite{shalf_2015}. 
Both conventional "More Moore" as well as newer "More than Moore" approaches in semiconductor design require highly precise, uniform, selective, and low-damage etching processes \cite{Oehrlein_2024,Lee_2014,Marchack_2021}. 
To meet these demands, precise control over the ion energy distribution function at the wafer is essential. 
 
One of the most frequently used reactor types for plasma etching are inductively coupled plasmas (ICP) with substrate bias, as they feature a high plasma density and thus high etch rates, as well as control over the ion energy \cite{Chabert2011, Smith_1997, Donelly_2013}. The plasma is generated by an RF-driven coil (source) and a second RF power supply is connected to the substrate to accelerate ions onto it (bias). 
This approach bears two inherent limitations: (1) The use of a sinusoidal RF bias results in a wide, bimodal IEDF, which contradicts the goal of precise ion energy tuning and control of selectivity \cite{Lieberman2005, Kawamura_1999}. 
For processes that require a certain threshold ion energy, such as atomic layer etching (ALE) \cite {kanarik2015} or sputter etching, ions with energies that exceed the threshold can compromise precision and selectivity or lead to plasma-induced damage of the substrate. Less energetic ions have insufficient energy to contribute to the process, resulting in energy waste and process inefficiency. Therefore, ideally the IEDF would be monoenergetic and all ions would arrive at the substrate at or above the threshold energy. As DC voltages cannot be used for dielectric substrates and classical RF biases result in broad IEDFs, this process requirement cannot be met with conventional approaches. (2) Despite the two independent power supplies, there are various coupling mechanisms between the power sources so that the ion flux at the substrate cannot be controlled completely independently of the ion energy \cite{Schulze_2012, Lee_2012, Berger_2017, Hebner_2000}.

One approach to increase process control involves pulsing of the RF generators. This can be done by either pulsing the source or bias power or both simultaneously. In this way, the ion-to-neutral flux ratio as well as the ion energy can be manipulated to a certain degree \cite{Agarwal_2009,Agarwal_2012,Banna_2012,Economou_2014}.
Although pulsing can provide additional control over several crucial process characteristics, it is clear that the complex, coupled, and oftentimes nonlinear interactions of control and process parameters necessitate a thorough process tuning. 

A different attempt to achieve not only independent control of the average ion energy but also of the shape of the IEDF is voltage waveform tailoring (VWT) for substrate biasing, i.e., the use of non-sinusoidal waveforms.
VWT has been implemented in capacitively coupled plasmas (CCPs) and was shown to be a powerful tool to tune many process characteristics, among them ion energy, radical density, and electron dynamics \cite{Lafleur2012, Johnson_2011, Derzsi_2013,lafleur2015tailored,Wang_2021,Hartmann_2021,Dong_2025,Dong_2025x}. 
In ICPs, the common RF sinusoidal bias voltage can be replaced by a tailored pulse-wave shaped voltage waveform with a short duty cycle. For the most part, the substrate voltage is at a constant negative value, which will accelerate the ions to a single energy in a collisionless sheath at low pressure. During the short positive voltage part, the electron current reaches the surface, compensating the positive surface charging of the dielectric wafer caused by ion bombardment during the remainder of the pulse period \cite{Schulze_2010x}. 
The group of Amy Wendt showed theoretically and experimentally that the generation of mono-energetic IEDFs in helicon plasma sources is possible using non-sinusoidal waveforms at a base frequency of \SI{500}{\kilo\hertz} for substrate biasing \cite{Wang_2000, patterson_2007,buzzi_2009, Qin_2010}. Faraz \etal \cite{faraz2020precise} used the same method at a lower frequency of \SI{100}{\kilo\hertz} to quantify the sputter yield of argon ions on a silicon dioxide (\sd{}) substrate. In their conclusion, they proposed the possibility of selective atomic-scale processes enabled only by tailored waveforms. Hartmann et al. \cite{Hartmann_2023} computationally demonstrated that such low-frequency tailored voltage waveforms lead to time-dependent positive surface charging of a dielectric substrate due to positive ion bombardment during most of the pulse period in the absence of any electron current, which only reaches the substrate during the short sheath collapse. This dynamic positive surface charging depends on the ion flux towards the wafer as well as its impedance, and thus on its material, thickness, and driving frequency. Such charging was demonstrated to cause shielding of a time-dependent fraction of the externally applied voltage from the plasma depending on the instantaneous surface charge on the wafer. This shielding effect affects the sheath voltage and can thus induce a parasitic broadening of the IEDF. To avoid such parasitic effects, instead of a rectangular driving voltage waveform with a flat plateau, a ramped voltage waveform was applied.

Previous work is restricted to limited sets of process conditions and covered only IEDF tuning as well as measurements of etch rates, without discussing other potential effects of the tailored bias voltage.

Therefore, in this work, we perform a multidiagnostic experimental characterization of such discharges within a large parameter range to highlight both the advantages and the limitations of tailored substrate bias voltages in a commercial inductively coupled reactive ion etching (RIE) tool. 
We investigate the influence of a tailored substrate bias voltage in ICPs on (1) the IEDF shape and ion flux, (2) electron dynamics and density, and (3) etch rates and selectivity. In section 2, the experimental setup is introduced, while results are presented in section 3, which is divided into three parts. In the first part, the effects of the substrate bias on the IEDF shape and ion flux are determined to demonstrate that monoenergetic IEDFs can be realized both in argon as well as in sulfur hexafluoride gas.  In the second part, phase-resolved optical emission spectroscopy (PROES) and a multipole resonance probe (MRP) are used to explore the effects of the substrate bias on electron dynamics and density. In the final part, the sputter threshold ion energies for \sd{} and amorphous silicon (a-Si) are determined based on monoenergetic IEDF generated via VWT, revealing an energy window where a-Si can be selectively etched against an \sd{} substrate by argon ion sputtering, if narrow IEDFs are ensured via VWT.

\section{Experimental setup}
\label{sec:methods}

\subsection{Plasma reactor}
\begin{figure}[!h]%here, top, bottom
        \centering
        \includegraphics[width=0.8\textwidth,trim=0 00 0 0, clip]{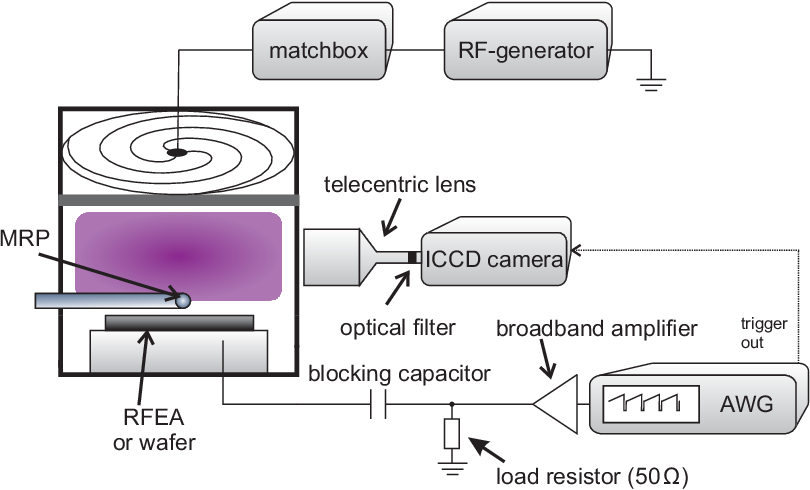}
        \caption{Schematic of the experimental setup including diagnostics. }
        \label{fig:schematic}
    \end{figure}
\begin{figure}[!h]%here, top, bottom
        \centering
        \includegraphics[width=0.8\textwidth,trim=30 00 10 0, clip]{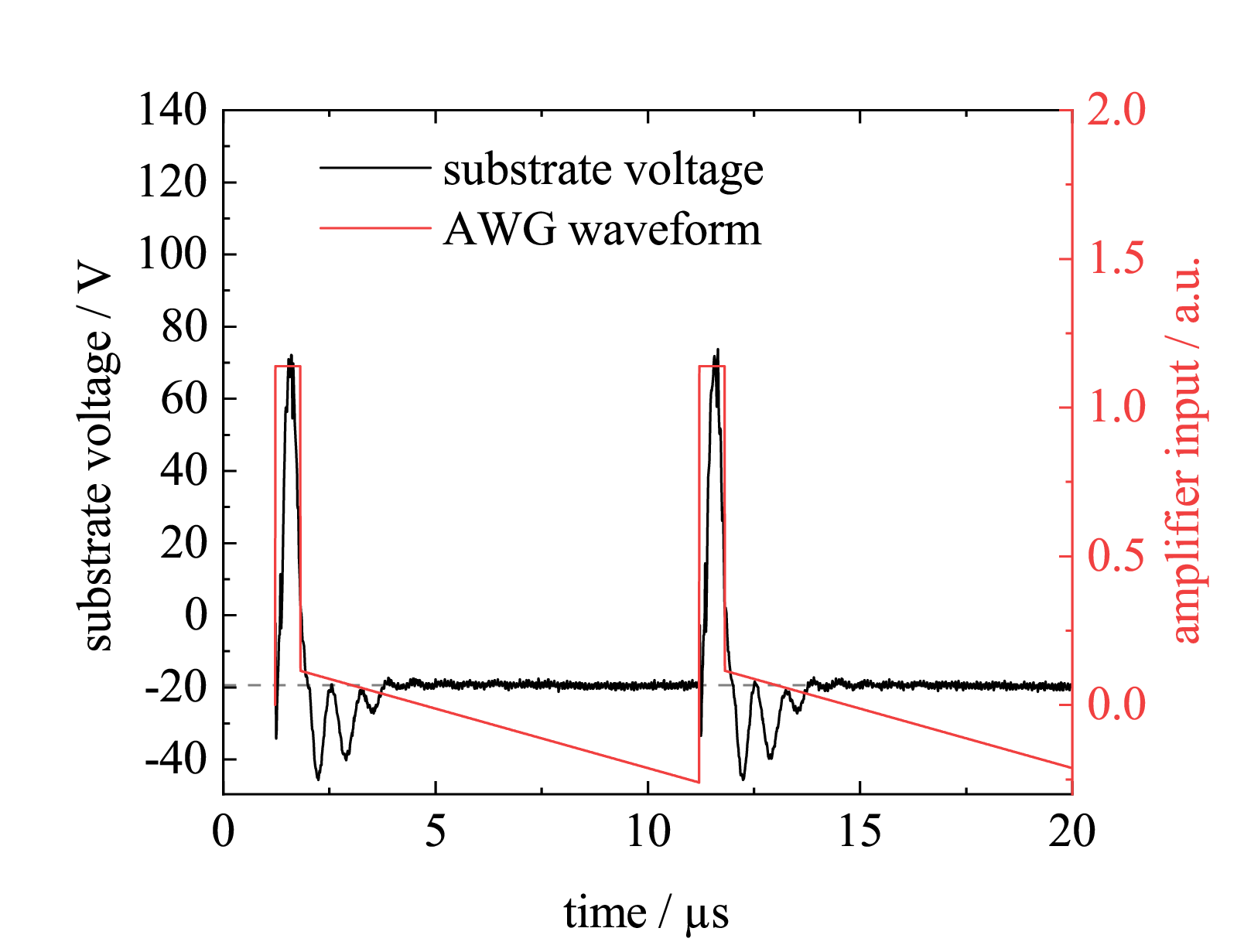}
        \caption{Exemplary tailored voltage waveform used for substrate biasing. The red line shows the signal of the AWG, while the black line shows the resulting substrate voltage measured by a high voltage probe at the substrate. The dashed line indicates the plateau voltage $V_{\rm p}$, which is the quasi-DC value of the negative plateau. }
        \label{fig:waveform}
\end{figure}

All measurements were carried out in a commerical SI 500 Reactive Ion Etching (RIE) reactor for \SI{200}{\milli\meter} wafers built by Sentech Instruments GmbH, schematically shown in figure \ref{fig:schematic}. 
The reactor is part of a six-port cluster tool with a wafer transfer system, which allows chaining of different processing steps without breaking vacuum.
The system uses a planar triple spiral antenna for the plasma generation with a \SI{13.56}{\mega\hertz} sine wave generator and matching system. 
The cylindrical vacuum chamber with a diameter of \SI{350}{\milli\meter} was expanded by a custom-made \SI{10}{\centi\meter} tall diagnostic ring, which provides additional flanges for diagnostic access (not shown in the sketch). With the diagnostic ring in place, the size of the gap between the ICP coil and the substrate is \SI{206}{\milli\meter}. 
The reactor is evacuated by a two-stage pumping system, reaching a base pressure below \SI{1e-5}{\pascal}. Process gases are supplied through an inlet near the ICP coil, and the operating pressure is set by a throttle valve and an automatic closed-loop control. For all measurements, the gas pressure was set to \SI{1}{\pascal} with a gas flow of 30\,sccm.
Wafers are placed on the substrate electrode and fixed by mechanical clamping with a weighted clamping ring. 
The substrate electrode can also be powered by a \SI{13.56}{\mega\hertz} sine wave. For the majority of this work, this system was replaced by a low frequency voltage waveform tailoring (LF VWT) system. This VWT system consists of an arbitrary waveform generator (AWG, here: 33600A, Keysight Technologies), a broadband power amplifier (VBA100-1100, Vectawave) with \SI{50}{\ohm} load resistor in parallel to the output and a \SI{1}{\micro\farad} blocking capacitor. The AWG generates a \SI{100}{\kilo\hertz} pulse wave signal with a \SI{15}{\percent} duty cycle ($D$) which is amplified by the power amplifier and then applied to the substrate through the blocking capacitor (see figure \ref{fig:waveform}). The load resistor is used to ensure a \SI{50}{\ohm} load which is required by the amplifier. During the phase of negative voltage, a linear voltage ramp with a negative slope is added to the signal to compensate for the dynamic charging effects of the blocking capacitor, similar to the approach of Hartmann et al. for compensating the effects of dynamics wafer charging \cite{Hartmann_2023}. In our work, only highly conductive silicon wafers with thin dielectric surface layers are placed on the substrate electrode. They feature a large capacitance and, thus, a small impedance so that dynamics shielding effects due to wafer charging are negligible. The slope of the bias voltage waveform is manually adjusted based on measurements of the voltage waveform at the substrate by a high voltage probe below the wafer to ensure a flat negative plateau of voltage waveform at the substrate. In the absence of additional dynamic shielding effects caused by wafer charging, this substrate voltage waveform corresponds to the voltage drop across the plasma and its flat negative plateau ensures the presence of a constant voltage drop across the sheath at the wafer for most of the pulse period. Figure \ref{fig:waveform} shows exemplary tailored voltage waveforms used for substrate biasing and generated by the AWG (red line), which includes the ramp, and measured at the substrate (black line). The oscillations that are visible in figure \ref{fig:waveform} between \SI{2}{\micro\second} and \SI{4}{\micro\second} are caused by an internal resonance of the amplifier due to the steep edges of the input voltage. Heavy ions hardly react to such oscillations due to their high inertia, so that their effects on the IEDF are negligible. 

The peak-to-peak value of the substrate voltage and, thus, the ion energy can be tuned by changing the peak-to-peak value of the voltage waveform generated by the AWG. The blocking capacitor used here limits the maximum peak-to-peak voltage on the substrate to approximately \SI{180}{\volt} for safety reasons. This leads to a maximum achievable ion energy of \SI{110}{\electronvolt} to \SI{130}{\electronvolt}, depending on the substrate material.

\subsection{Diagnostics}
To measure the ion flux and IEDF at the substrate, a retarding field energy analyzer array (RFEA, here: Semion RFEA, Impedans LTD) was used. The array consists of nine sensors that are mounted on a circular plate with a diameter of \SI{200}{\milli\meter}. The plate is made of anodized aluminum and is placed on the substrate electrode, replacing the wafer. For each IEDF measurement, five measurements are averaged. A maximum energy resolution of \SI{1}{\electronvolt} is used. The system is not absolutely calibrated, so that all measured IEDFs will be given in arbitrary units (\si{\au}) \cite{Ries2021}. In this work, only the data measured by the central sensor is used, i.e., measurements are performed at the substrate center.

Phase-resolved optical emission spectroscopy (PROES) is used to measure the influence of the tailored substrate bias on the electron dynamics. The fundamentals of this diagnostic can be found in \cite{Schulze_2010}. We investigate the electron impact excitation rate from the ground state into the Ar 2p\textsubscript{1} level, which has an electron impact excitation threshold energy of \SI{13.5}{\electronvolt} \cite{NIST_ASD}. An intensified charge-coupled device camera (ICCD camera, here: 4Quick, Stanford Computer Optics) in combination with a telecentric lens (TOB42/11.0-185-V-WN, Vision \& Control GmbH) and an optical bandpass filter with a wavelength of \SI{750}{\nano\meter} (FWHM \SI{10}{\nano\meter}) is focused on the plasma directly above the substrate electrode. The image acquisition is triggered by the AWG with a variable delay and a gate time of \SI{100}{\nano\second}. As the ICP driving voltage waveform and the subtrate bias voltage waveform are not phase-locked, temporal resolution is realized only with respect to the electron dynamics caused by the substrate bias. Excitation caused by electron acceleration by the azimuthal electric field induced by the ICP coupling appears as a constant background signal, similar to the results reported in \cite{Schulze_2012}. Each spatio-temporally resolved plot of this electron impact excitation rate ranges over a span of \SI{12}{\micro\second} which equals \num{1.2} periods of the substrate voltage. The substrate voltage waveform is shown below each spatio-temporally resolved plot of the excitation rate. 

To measure the electron density, a multipole resonance probe (MRP, House of Plasma GmbH) is used, which is an active plasma resonance spectroscopy diagnostic. The details of the working principle can be found in \cite{Lapke2008, Lapke2011, Oberrath2021, Schulz2014, Fiebrandt2017}. The MRP couples a high-frequency signal in the range of \SI{100}{\mega\hertz} to \SI{10}{\giga\hertz} into the plasma and measures its absorption. From the absorption spectra, the electron density and electron temperature are calculated using a mathematical model. The probe is encapsulated in a glass rod with a diameter of \SI{8}{\milli\meter}. The probe head is placed in the center of the discharge, \SI{1}{\centi\meter} above the substrate. Both the MRP and the RFEA array are only inside the chamber during their respective measurements. When the RFEA array is not in use, an n\textsuperscript{+}-doped silicon wafer is placed on the substrate electrode. 

For the characterization of the etch rates, commercially available degenerately doped silicon wafers with a \SI{60}{\nano\meter} thick thermal oxide layer were used. These wafers were either used out of the box, or an approximately \SI{30}{\nano\meter} thick layer of aluminum oxide (Al\textsubscript{2}O\textsubscript{3}) or amorphous silicon (a-Si) was deposited on the surface. The Al\textsubscript{2}O\textsubscript{3} wafer was prepared using a plasma-enhanced atomic layer deposition process \cite{Willeke_2022}. The a-Si wafer was deposited via capacitively coupled plasma-enhanced chemical vapor deposition using \SI{2}{\percent} silane as the precursor in an Oxford Plasma Pro 100 reactor. 

The etch depths of \sd{} and Al\textsubscript{2}O\textsubscript{3} were measured using a spectroscopic ellipsometer (SE800 DUV, Sentech Instruments GmbH) under atmosperic conditions. The etch depth of a-Si was measured using a laser ellipsometer (Al Real Time Monitor, Sentech Instruments GmbH) that is mounted in a different processing chamber of the cluster tool. This allows measurements of the silicon layer without exposing the sample wafer to air, and thus the formation of a native oxide layer is suppressed. For each data point, the remaining layer thickness was measured before and after plasma exposure by the respective ellipsometer, and the difference was used to calculate the etch rate.

\section{Results}
\subsection{Tailored ion energy distribution functions}
\label{sec:IEDF}
In the following sections, measurements of the IEDF at the substrate electrode are presented in the presence of different bias voltage waveforms and different ICP source powers.
\subsubsection{RF bias\\}
\begin{figure}[!h]%here, top, bottom
        \centering
        \begin{subfigure}{0.47\textwidth}
        \includegraphics[width=\textwidth,trim=00 0 0 20, clip]{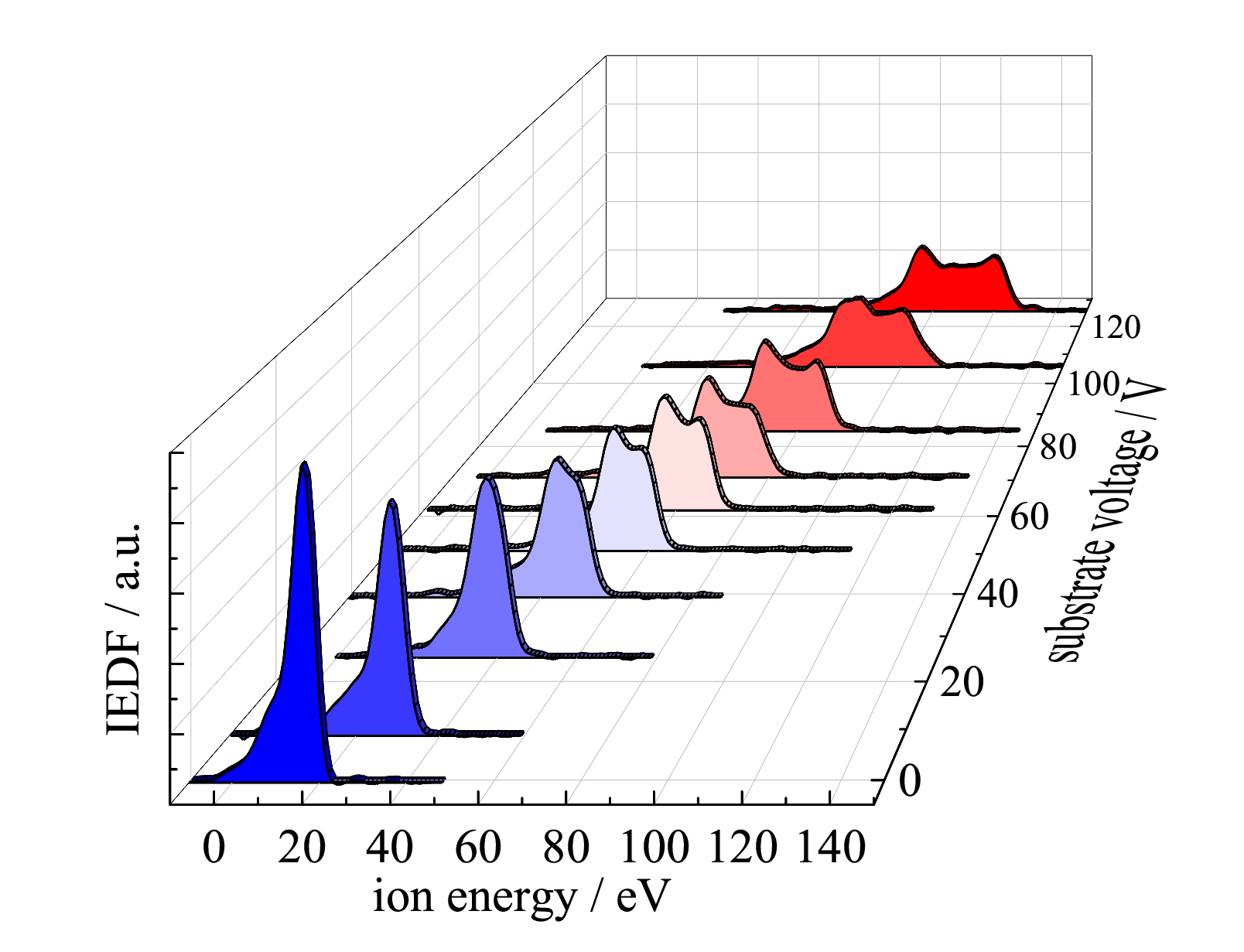}
        \caption{\SI{100}{\watt} ICP source power}
        \label{fig:100WRF}
        \end{subfigure}
        \hspace*{\fill}
        \begin{subfigure}{0.47\textwidth}
        \includegraphics[width=\textwidth,trim=00 0 0 20, clip]{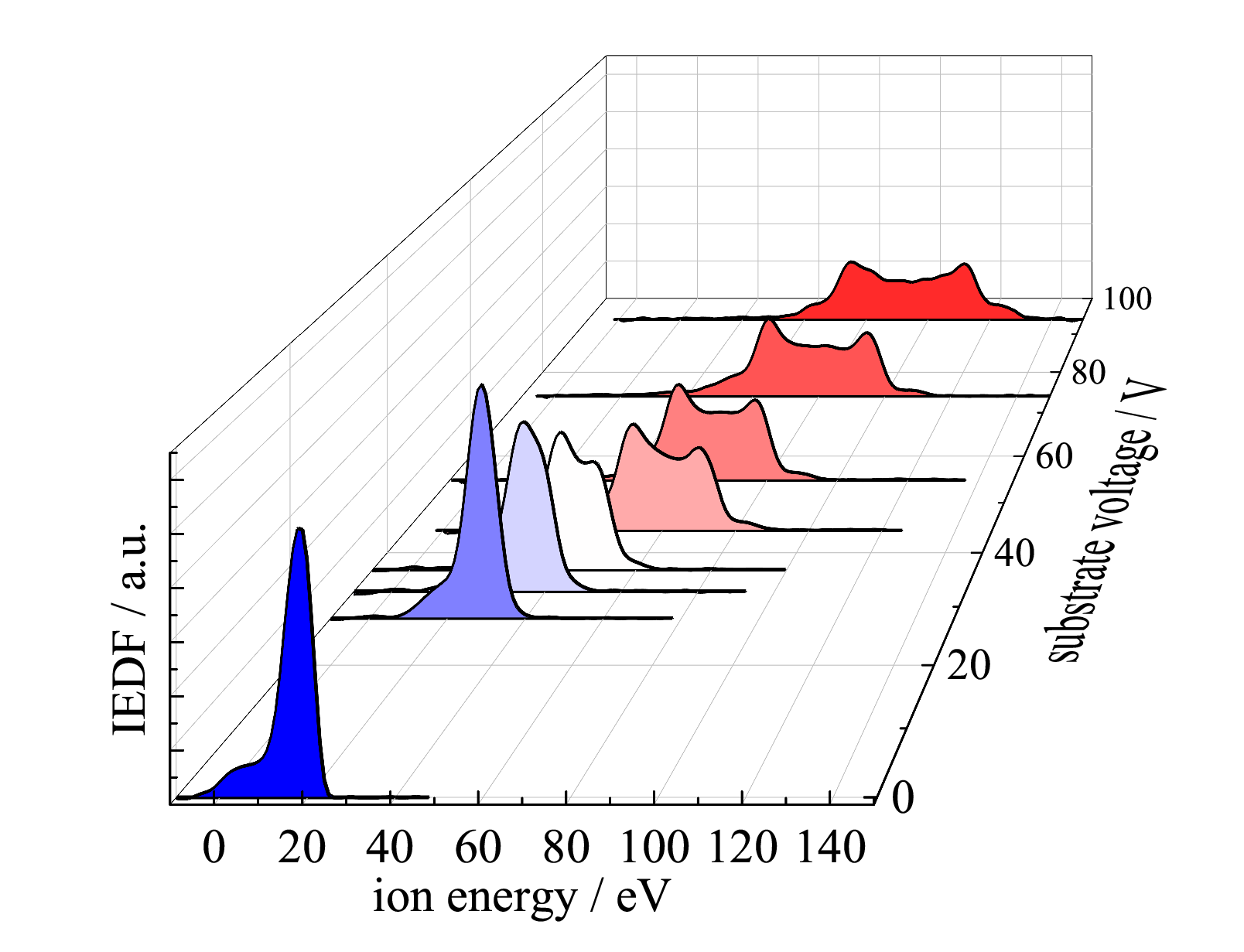}
            \caption{\SI{300}{\watt} ICP source power}
            \label{fig:300WRF}   
        \end{subfigure}
        \hspace*{\fill}
        
        \begin{subfigure}{0.47\textwidth}
        \includegraphics[width=\textwidth,trim=00 0 0 20, clip]{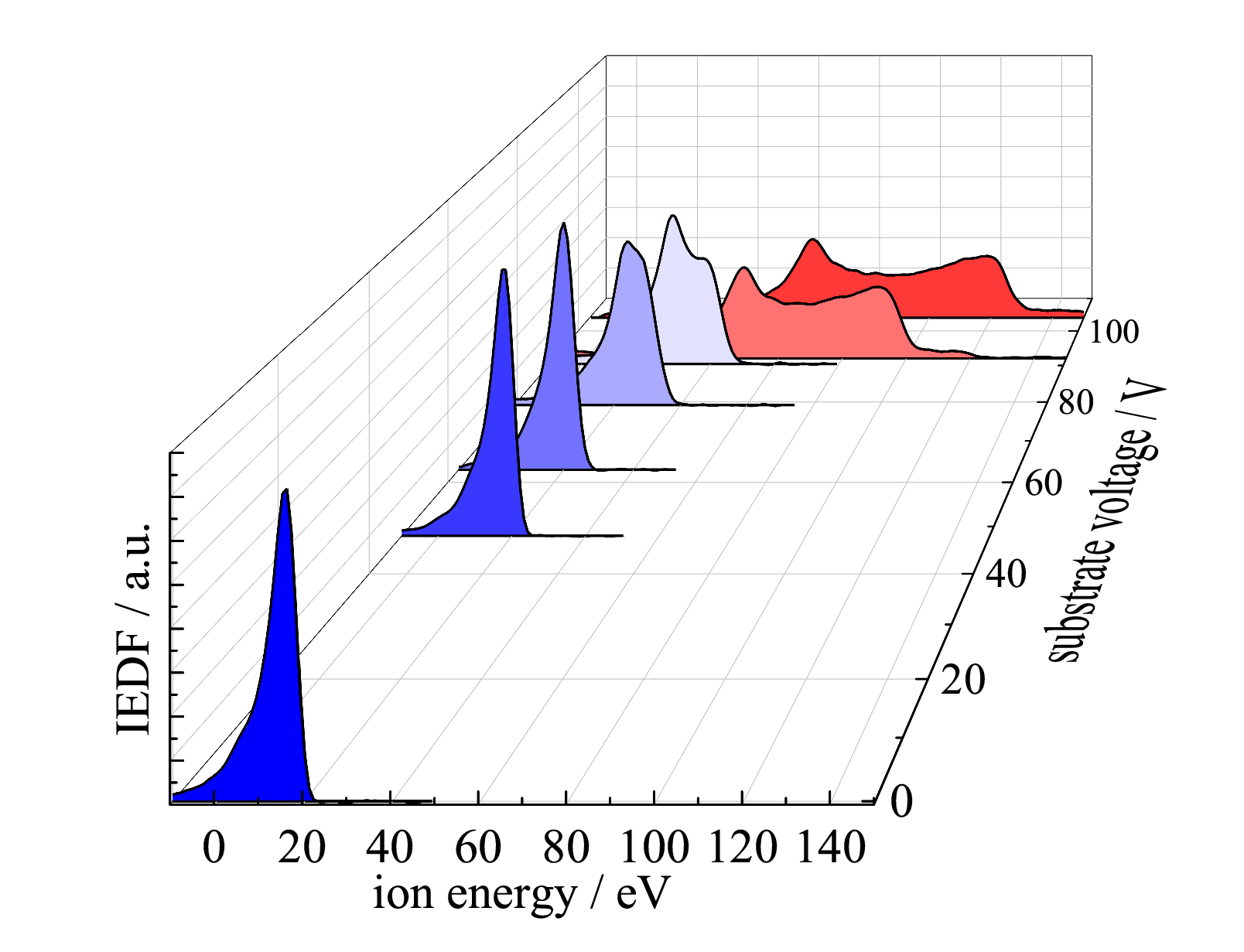}
            \caption{\SI{700}{\watt} ICP source power}
            \label{fig:700WRF}   
        \end{subfigure}
        \hspace*{\fill}

        \caption{IEDFs at the substrate electrode measured in argon at \SIlist{100;300;700}{\watt} ICP source power at \SI{1}{\pascal} using the sinusoidal RF-bias (13.56 MHz). The substrate voltage is measured as peak-to-peak value.}
        \label{fig:RFBias}
\end{figure}
First, the conventional sinusoidal RF bias (13.56 MHz) is used and the resulting IEDFs are measured as a function of the peak-to-peak voltage of the RF bias for different ICP source powers in argon gas at 1 Pa. Figure \ref{fig:RFBias} shows the corresponding IEDF measurements at \SIlist{100;300;700}{\watt} ICP source power. 

For 0 V substrate voltage, the substrate is floating because of the presence of the blocking capacitor between the substrate and ground. This results in a narrow monoenergetic IEDF with an energy peak at around \SI{17}{\electronvolt} and a full width at half maximum (FWHM) of \SI{6.1}{\electronvolt}. Assuming a collisionless sheath, the peak ion energy is determined by the sum of the presheath potential and the floating potential. This leads to a peak ion energy of 
\begin{equation}
    E_i =\frac{T_e}{2}\left(1 + \mathrm{ln}\left( \frac{m_i}{2\pi m_e} \right) \right) \, , 
    \label{eq:Eif}
\end{equation}
where $T_e$ is the electron temperature in \si{\electronvolt}, $m_i$ the ion mass and $m_e$ the electron mass \cite{Lieberman2005}. The fact that the peak ion energy at this floating surface is constant at different source powers shows that $T_e$ is not affected by the source power. According to equation (\ref{eq:Eif}), $T_e$ is \SI{3.2}{\electronvolt} based on the measured ion energy.\\
With the bias voltage turned on, the shape of the IEDF transitions into the well known bimodal structure \cite{Lieberman2005,Kawamura_1999}. The separation between the two peaks along the energy axis increases both with increasing the ICP source and bias power and can reach up to \SI{70}{\electronvolt} at \SI{700}{\watt} of ICP source power with a bias voltage of \SI{104}{\volt} peak-to-peak. The observed increase of the IEDF width as a function of power is caused by the corresponding increase of the ion density and, thus, of the ion plasma frequency. At high powers and high plasma density, the ions can more easily follow the instantaneous sheath electric field and, therefore, the IEDF gets broader. Such broad IEDFs are detrimental for etch selectivity and control.

\subsubsection{Low frequency VWT  bias at low ICP source power\\}

\begin{figure}[!h]%here, top, bottom
        \centering
        \includegraphics[width=0.8\textwidth,trim=00 00 0 0, clip]{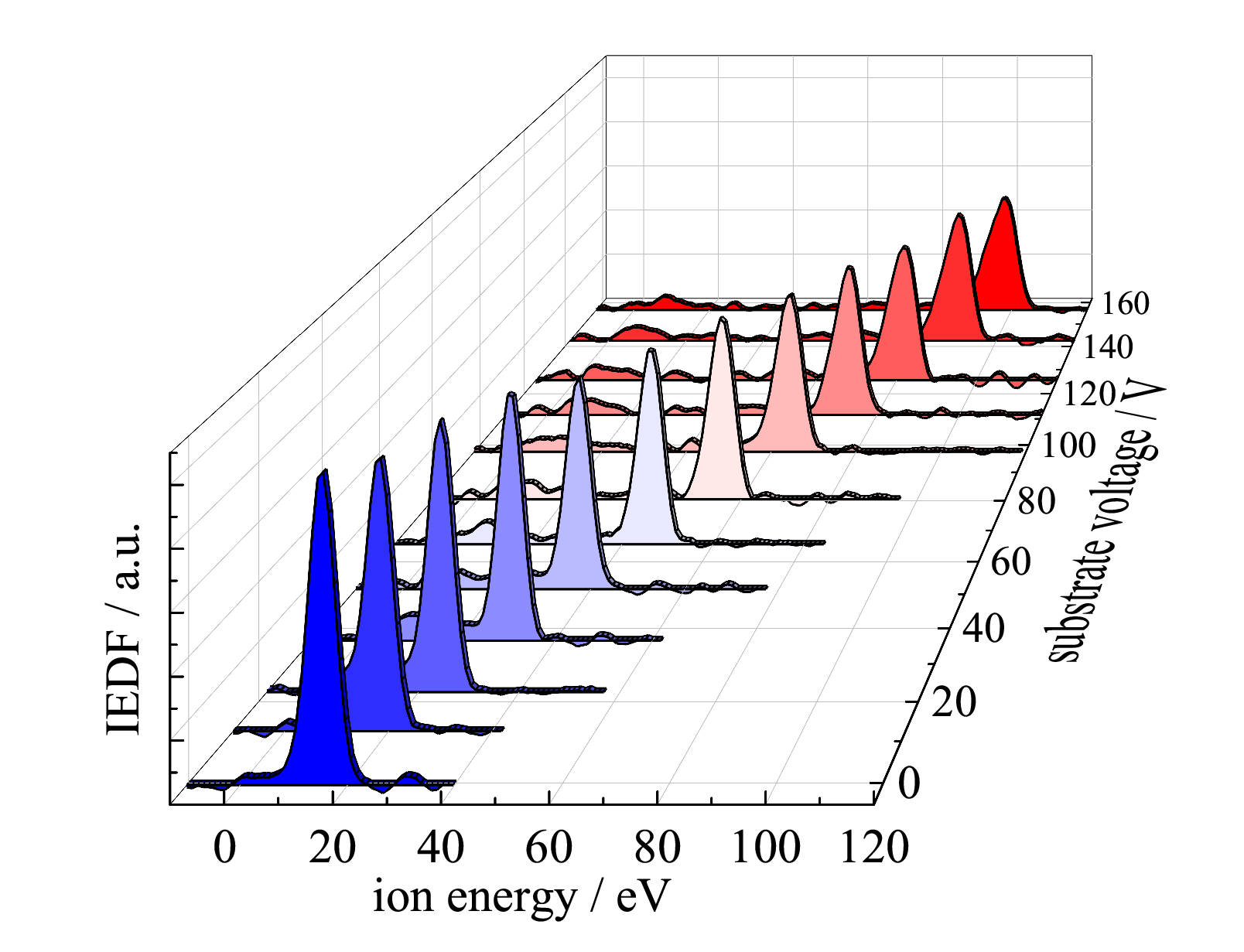}
        \caption{IEDFs at the substrate electrode measured in argon at \SI{100}{\watt} ICP source power and varying LF VWT substrate voltages at \SI{1}{\pascal}. The substrate voltage is measured as peak-to-peak value.}
        \label{fig:IEDFAr100W}
\end{figure}

For the remainder of this work, the single frequency RF bias generator and matchbox, originally connected to the substrate electrode, were removed and replaced by the VWT system described in section 2.1.
Figure \ref{fig:IEDFAr100W} shows the IEDFs generated by the low-frequency pulse-wave bias with \SI{100}{\watt} ICP source power. In contrast to the results obtained for the classical single frequency RF bias, the narrow single-peak structure is maintained throughout the entire parameter range. Because the duration of the negative voltage plateau is much longer than the ion transit time through the sheath, ions are accelerated by a quasi-DC electric field in a collisionless sheath, leading to the narrow peak of the IEDF. Since the duty cycle D (\SI{15}{\percent}) of the bias waveform is small, the majority of ions is accelerated towards the electrode during the time of the negative plateau and the effect of the positive part of the waveform is barely visible in the IEDF measurements. At higher substrate voltages, a small peak is visible at low ion energies, which is most likely caused by the positive part of the substrate bias voltage waveform.
The FWHM of the peaks increases marginally with increasing bias voltage from \SI{5,95}{\electronvolt} to \SI{9,34}{\electronvolt}. The ion flux remains nearly constant with a deviation of \SI{2.75}{\percent}, aligning with the goal of independent IEDF control. Such a narrow energy spread realized via VWT makes these IEDFs attractive for selective ALE applications \cite{faraz2020precise} and other plasma processes.

%-----------

\begin{figure}[!h]%here, top, bottom
        \centering
        \includegraphics[width=0.8\textwidth,trim=0 00 0 0, clip]{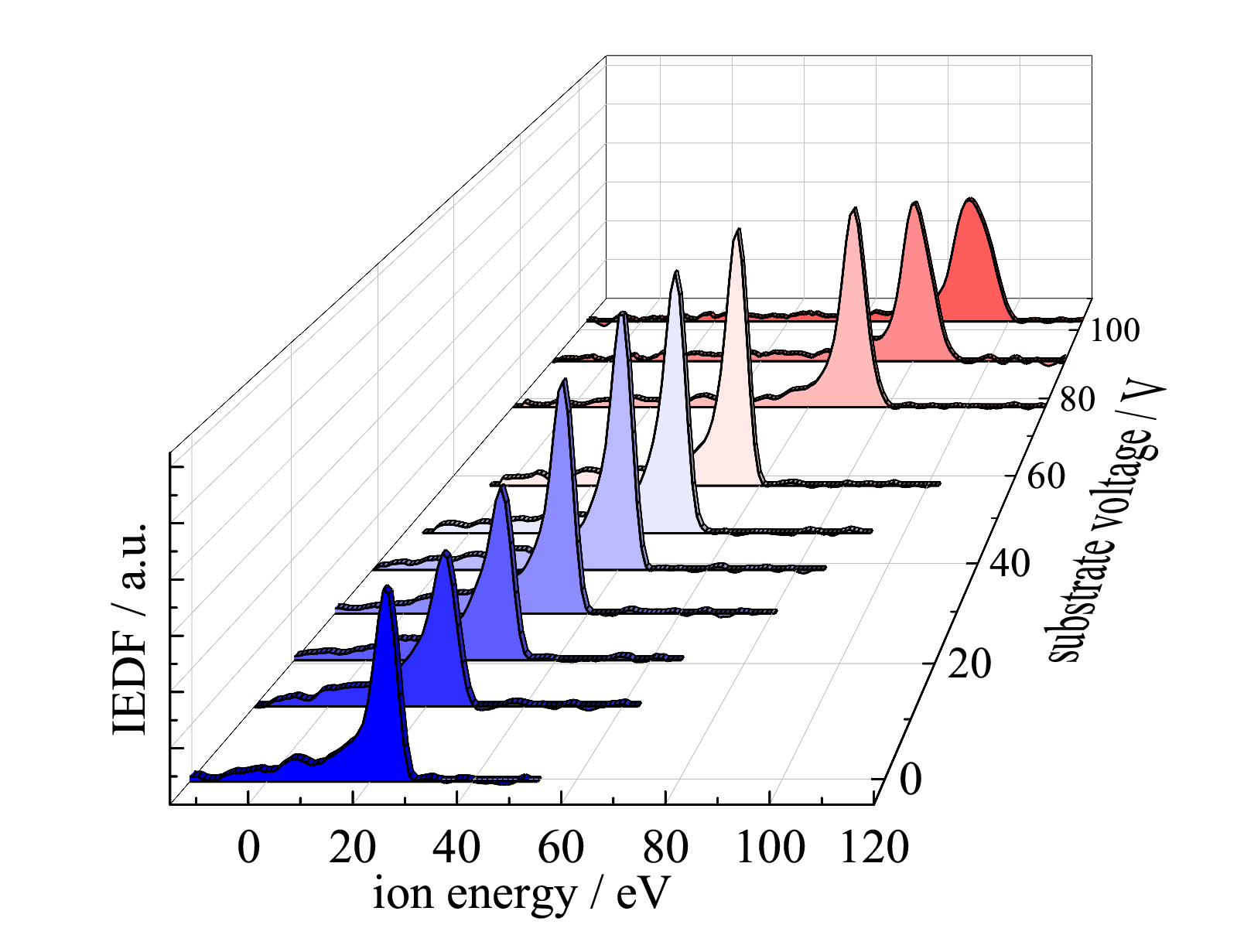}
        \caption{IEDFs at the substrate electrode measured in SF\textsubscript{6} at \SI{600}{\watt} ICP source power and varying LF VWT substrate voltages at \SI{1}{\pascal}. The substrate voltage is measured as peak-to-peak value.}
        \label{fig:IEDFSF6}
\end{figure}

The IEDF was also measured in the process-relevant gas sulfur hexafluoride (SF\textsubscript{6}) \cite{Agostino1981, Chang_2005}. Because SF\textsubscript{6} is a strongly electronegative gas \cite{Christophorou1995}, compared to argon discharges, higher ICP source powers are required to generate a stable discharge in the H-mode. For this setup, \SI{600}{\watt} was found to be the minimum ICP source power required for a stable H-mode in SF$_6$. The resulting IEDFs are shown in figure \ref{fig:IEDFSF6}. The measured IEDFs have a shape similar to that in the argon case with a FWHM between \SIlist{5.4;11.8}{\electronvolt}, but with a more populated low energy tail. This indicates the presence of a more collisional sheath compared to the argon discharge, potentially due to the lower positive ion density and, thus, larger sheath at fixed substrate voltage. 
These results show that VWT substrate biasing allows generating narrow IEDFs also in reactive/electronegative processing gasses.

To monitor the peak ion energy during actual etching processes, a measurement of thereof is needed, which does not require the presence of an RFEA sensor array inside the reactor. As the peak ion energy is determined by the quasi-DC negative plateau value of the substrate voltage waveform, $V_{\rm p}$ (see figure \ref{fig:waveform}), a measurement thereof is sufficient to determine the peak ion energy. In fact, a plot of the peak ion energy against this plateau voltage shows a linear relationship with a slope of \SI{-1}{\electronvolt\per\volt} and an offset of \SI{13}{\electronvolt}. These values were found to be constant at all ICP power levels. Thus, the peak ion energy can be calculated according to
\begin{equation}
    E_i = \mathrm{e} \left( \Phi_{\rm p} - V_{\rm p} \right), 
    \label{eq:Ei}
\end{equation}
where $\mathrm{e}$ is the elementary charge and $\Phi_{\rm p}$ is the offset, which is assumed to correspond to the floating potential \cite{Yu_2022}. Based on this equation, a closed-loop control was implemented, which keeps the plateau voltage at a constant level by changing the AWG output voltage, to ensure a constant peak ion energy even under drifting process conditions. This control was used for the etching experiments discussed in section \ref{sec:etchting}.

\subsubsection{Low frequency VWT bias at high ICP source power\\}

\noindent In this section, the LF VWT bias is applied at higher ICP source power levels.  

\begin{figure}[!h]%here, top, bottom
        \centering
        \includegraphics[width=0.8\textwidth,trim=30 00 40 0, clip]{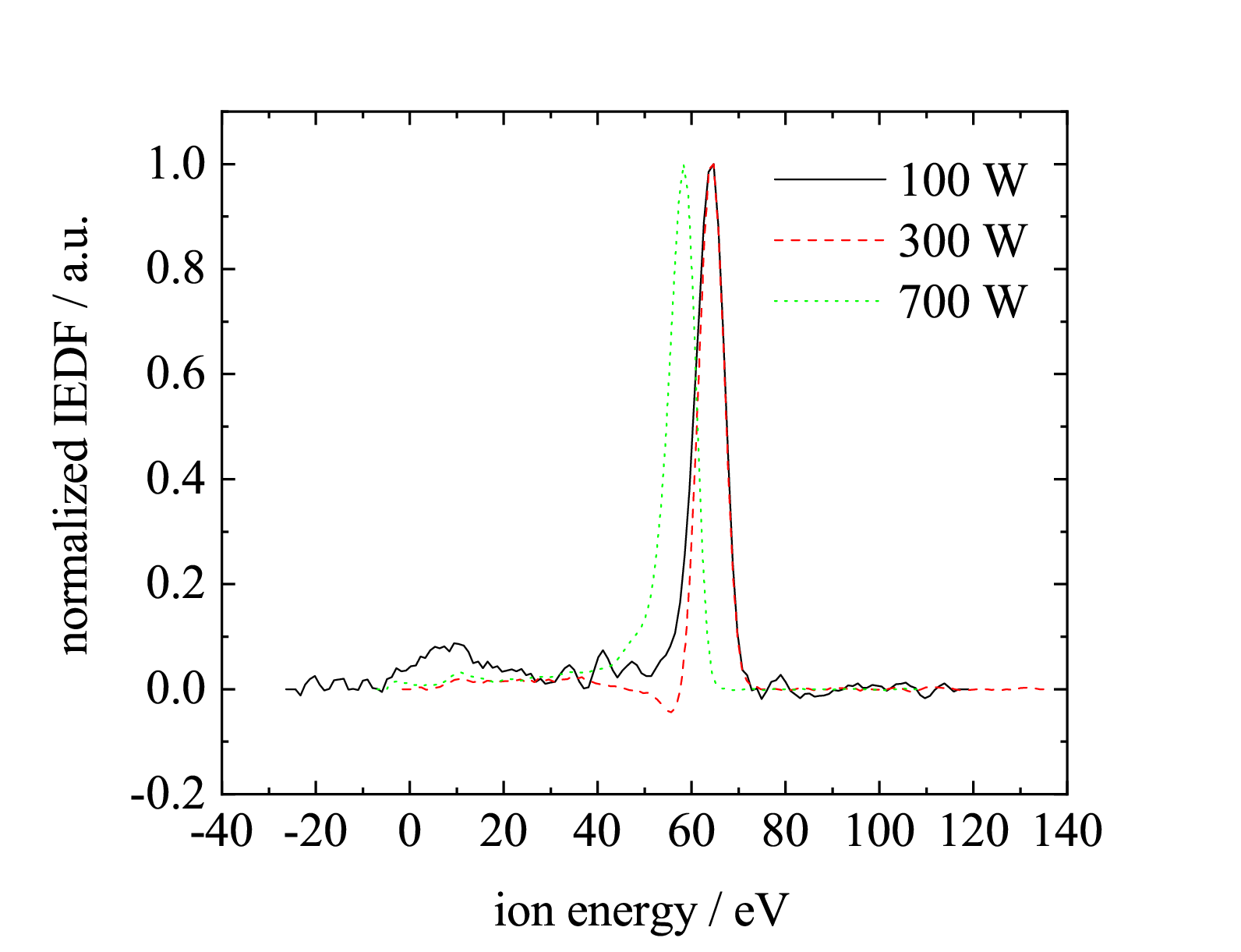}
        \caption{IEDFs at the substrate electrode measured in argon at \SI{1}{\pascal} and \SIlist{100;300;700}{\watt} ICP source power. The low frequency VWT bias is used and set to result in an IEDF peak at around \SI{60}{\electronvolt}. The IEDFs are normalized to their respective maximum.}
        \label{fig:normIEDF}
\end{figure}

\noindent In figure \ref{fig:normIEDF} three exemplary IEDFs measured at the substrate electrode at \SIlist{100;300;700}{\watt} ICP source power are shown. The measurements are selected from their respective measurement series to have an ion energy of approximately 60 eV.

As the ion fluxes are strongly different at the three ICP power levels, the data were normalized by their respective maximum values. The plot shows that the narrow shape of the IEDF is maintained, even at higher power levels, in strong contrast to the classical single frequency RF bias (see figure \ref{fig:RFBias}). 

\begin{figure}[!h]%here, top, bottom
        \centering
        \includegraphics[width=0.8\textwidth,trim=40 00 40 0, clip]{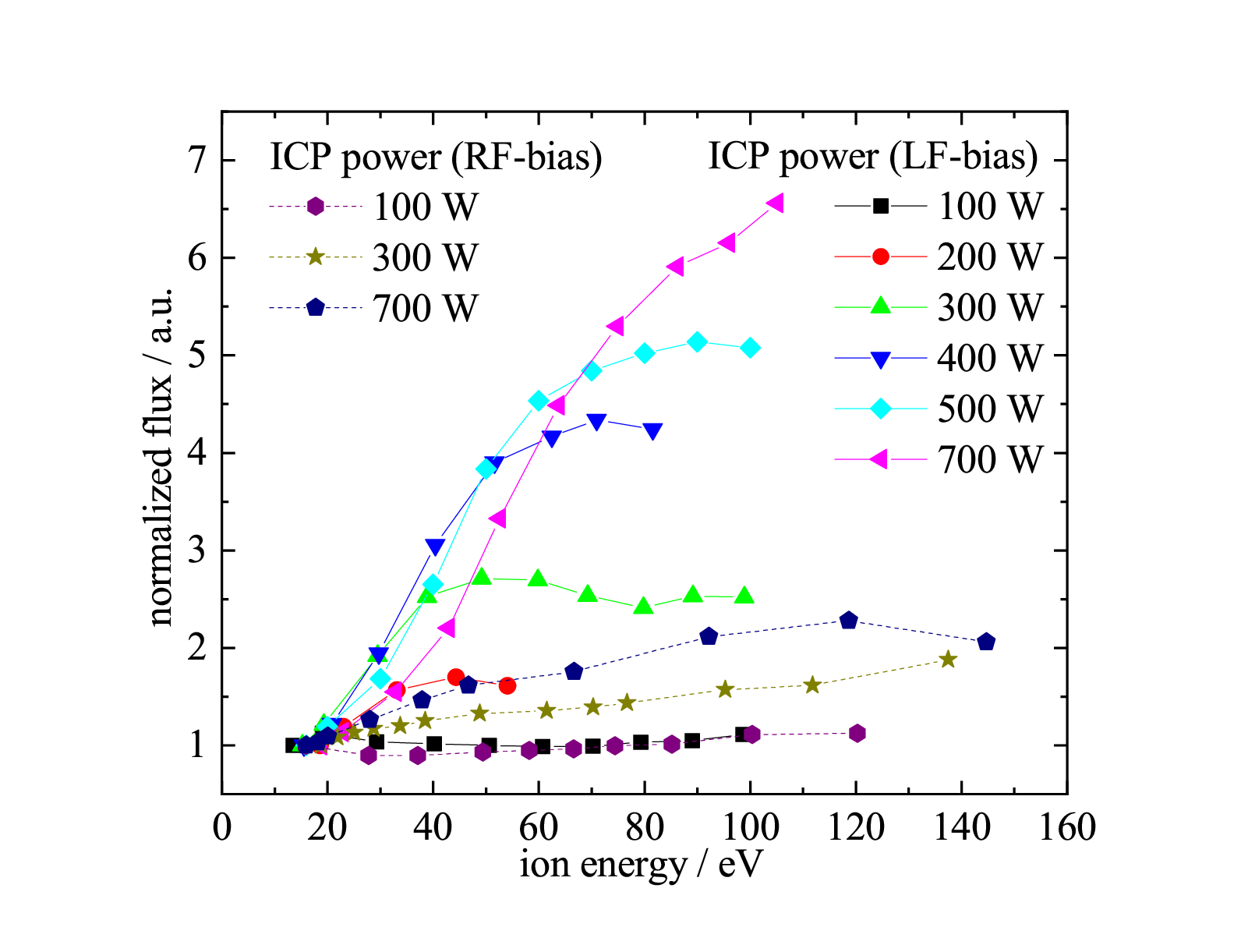}
        \caption{Ion fluxes at the substrate electrode measured in argon at \SI{1}{\pascal} at different ICP source powers using LF VWT (solid lines) and RF (dashed lines) substrate biases. The data are normalized by dividing each measured ion flux by the corresponding flux measured in the absence of any substrate bias. The ion energy corresponds to the position of the high energy IEDF peak on the energy axis and is controlled by adjusting the peak-to-peak voltage of the substrate bias voltage waveform and measured by an RFEA. }
        \label{fig:saturationPlot}
\end{figure}

At higher ICP source powers, the ion flux is no longer unaffected by the substrate bias. To depict this, in figure \ref{fig:saturationPlot} normalized measurements of the ion flux at the substrate electrode are shown as a function of the energy, at which the high energy peak of the IEDF occurs, at different ICP source powers and in the presence of either a classical single frequency RF bias or the low frequency VWT bias. The normalization is done by dividing each measured ion flux by the corresponding flux measured in the absence of any substrate bias. In this way, the vertical axis shows the relative increase of the ion flux caused by the bias voltage. The ion energy is controlled by adjusting the peak-to-peak voltage of the substrate bias voltage waveform.

At \SI{100}{\watt} ICP source power, the ion flux remains constant for both the LF VWT and the RF bias. In the RF bias case, at \SIlist{300;700}{\watt} ICP source power the ion flux increases moderately up to \num{2.3} times its initial value in the \SI{700}{\watt} case. 
For the LF VWT bias, at \SI{300}{\watt} ICP source power the ion flux first increases steeply up to \num{2.7} times the initial flux at an ion energy of \SI{50}{\electronvolt}. Thereafter, the ion flux saturates and remains approximately constant. At \SI{700}{\watt} ICP source power, the ion flux increases throughout the entire parameter range, reaching an increase of \num{6.5} of the value measured in the absence of any bias voltage. The data sets at \SIlist{200;400;500}{\watt} ICP source power show a similar behaviour as the \SI{300}{\watt} case. First, the ion flux increases by a factor of \num{1.7}, \num{4.3} and \num{5.1}, respectively, at an ion energy of \SI{44}{\electronvolt}, \SI{71}{\electronvolt} and \SI{90}{\electronvolt}, respectively. Then, the ion flux remains constant. These data sets were taken to confirm this saturation behavior and stopped after saturation became visible. Both the peak ion energy and, thus, the substrate peak-to-peak voltage at which the ion flux saturates, as well as the resulting ion flux, increase with the ICP source power. 

At constant plasma density in the plasma bulk (see figure \ref{fig:MRP}), the increase of the ion flux as a function of the LF VWT bias peak-to-peak voltage might be caused by a time modulation of the ion density at the substrate caused by the low repetition rate of the LF VWT voltage waveform, similar to results reported previously by Schulze et al. \cite{Schulze_2009}: During the sheath collapse phase at the substrate electrode (positive plateau of the substrate bias waveform), ions are not accelerated towards the electrode by a sheath electric field and, thus, the local ion density is high. When the sheath expands, the ions in vicinity of the electrode are accelerated towards this boundary surface and the local ion density and the ion flux decrease as a function of time. The higher the sheath voltage, i.e., the peak-to-peak voltage of the substrate bias as well as the peak ion energy, the larger the sheath and the more ions are accelerated towards the electrode, so that the ion flux increases as a function of the peak ion energy, controlled by the substrate bias voltage. At some point, however, the ion density is so low that the ion flux towards the electrode cannot be increased by increasing the substrat voltage anymore. At higher ICP source powers the ion density is higher in the plasma bulk and more ions flow into the sheath from that region so that it takes a higher substrate voltage to deplete the ion density in the sheath region at the wafer and the ion flux can raise to higher values. Ultimately, this phenomenon should be investigated by simulations to reveal a complete understanding of it.

\subsubsection{Triple-plateau VWT substrate bias\\}

\begin{figure}[h]%here, top, bottom
        \centering
        \includegraphics[width=0.8\textwidth,trim=40 00 40 0, clip]{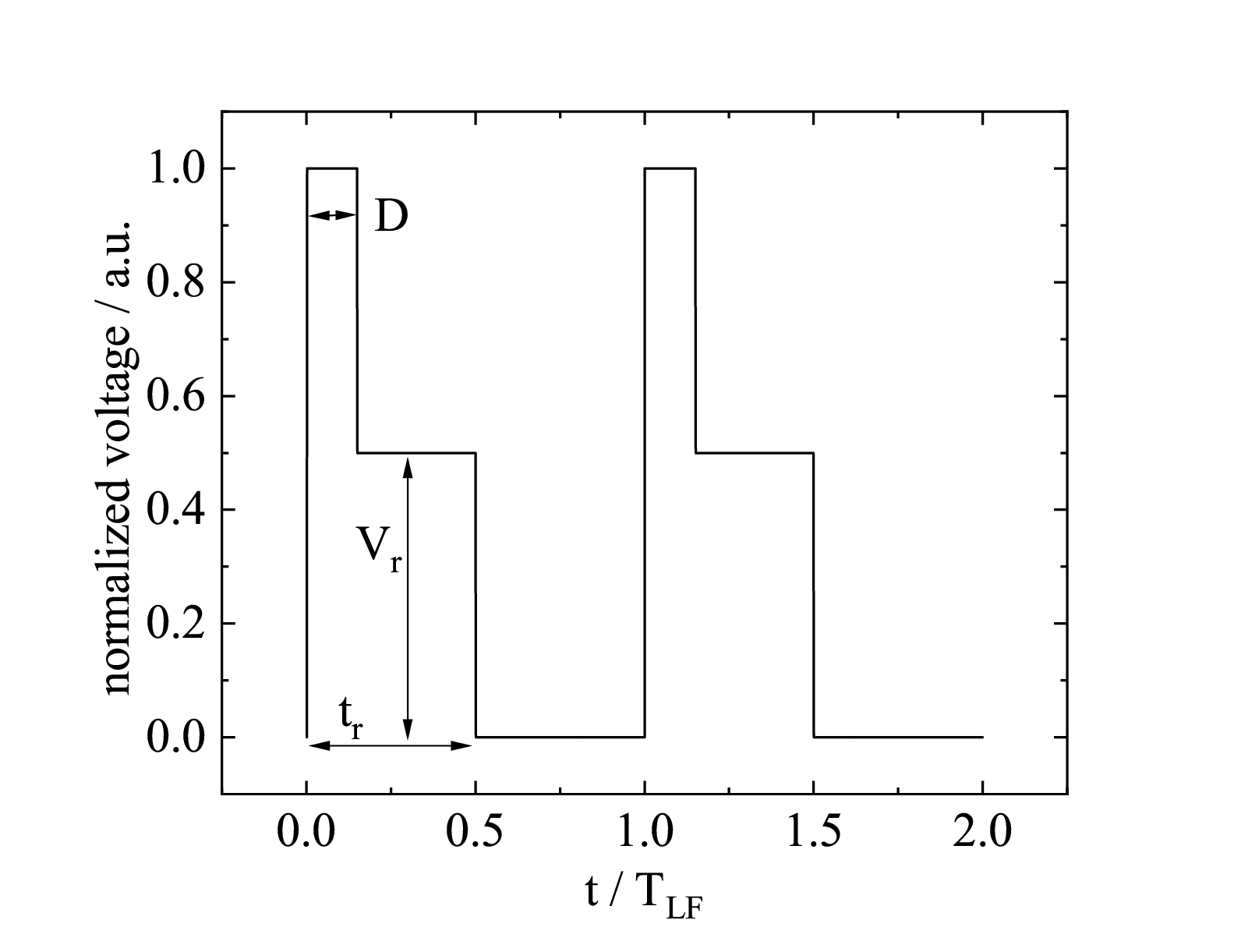}
        \caption{Schematic illustration of the triple-plateau VWT substrate bias voltage waveform used to generate an IEDF at the substrate with two fully controllable peaks. The arrows indicate the control parameters for the adjustment of these two IEDF peaks. }
        \label{fig:waveformDual}
\end{figure}

The IEDF shape can be further tailored, beyond generating a single narrow high energy peak, by using other waveform shapes for the substrate bias voltage waveform. For instance, a second peak can be added to the IEDF using a triple-plateau waveform shape, as shown in figure \ref{fig:waveformDual}. Such a waveform will lead to three different phases of sheath characteristics during one pulse period: (i) Sheath collapse (maximum positive voltage), (ii) Middle plateau determining the low energy IEDF peak, and (iii) lower plateau determining the high energy IEDF peak. Here, two additional voltage waveform control parameters are introduced: (i) The relative voltage ($V_r$) of the middle plateau compared to the lower plateau (see vertical arrow) controls the energy gap between the two IEDF peaks. A lower value causes the two plateaus to be closer to each other and, thus, the instantaneous sheath voltages during the corresponsing pulse fraction to be more similar so that the energy gap between the two IEDF peaks decreases. (ii) The relative on-time ($t_r$) of the middle plateau (horizontal arrow) controls the ratio between the ion flux in the respective IEDF peaks. A higher value yields a more pronounced low energy peak. The duty cycle $D$ was kept constant at \SI{15}{\percent} and a frequency of \SI{100}{\kilo\hertz} was used, resulting in a period $T_{LF}$ of \SI{10}{\micro\second}. 

\begin{figure}[!h]%here, top, bottom
        \centering
        \begin{subfigure}{0.49\textwidth}
        \includegraphics[width=\textwidth,trim=0 00 0 0, clip]{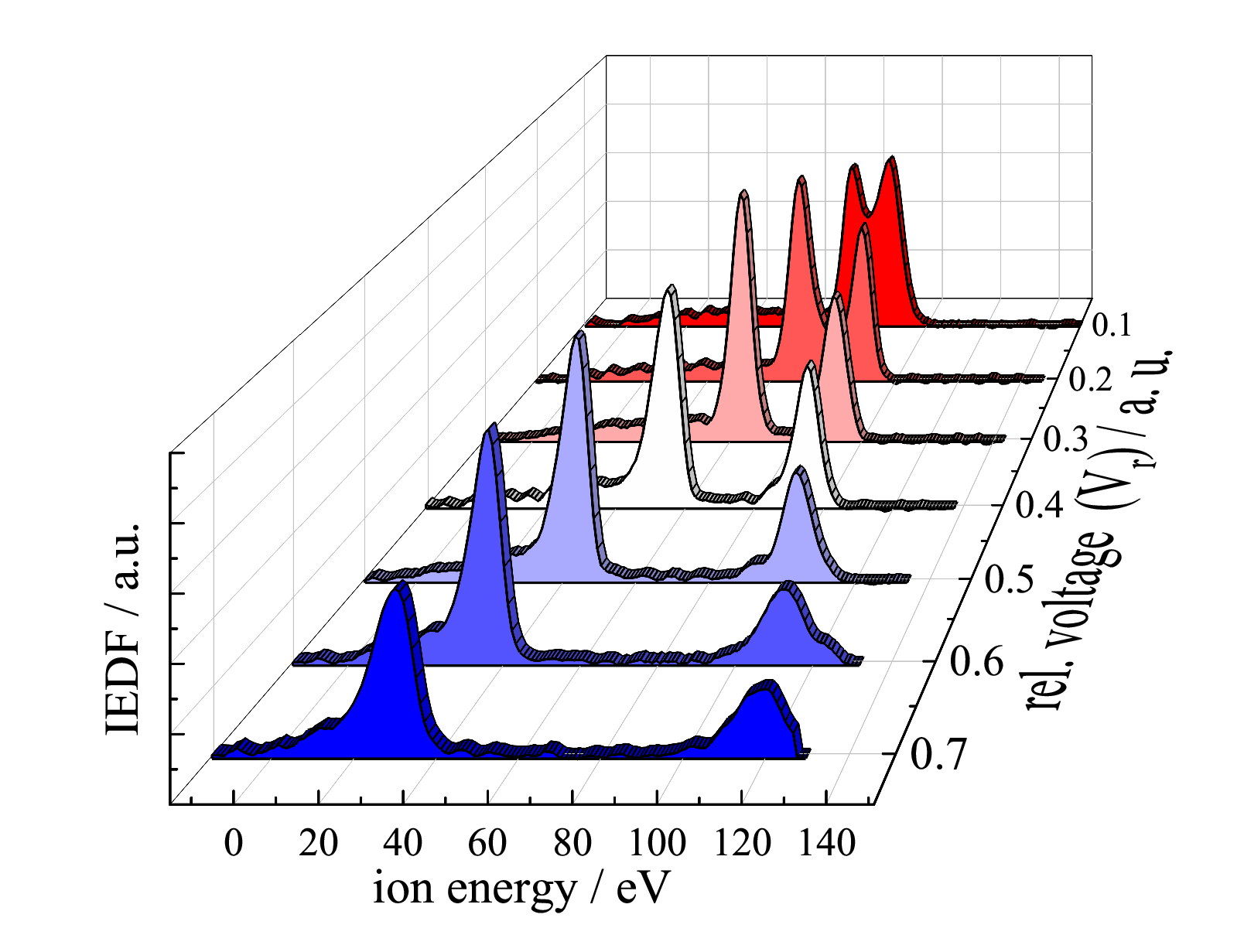}
        \caption{}
        \label{fig:relV}
        \end{subfigure}
        \hspace*{\fill}
        \begin{subfigure}{0.49\textwidth}
        \includegraphics[width=\textwidth,trim=0 00 0 0, clip]{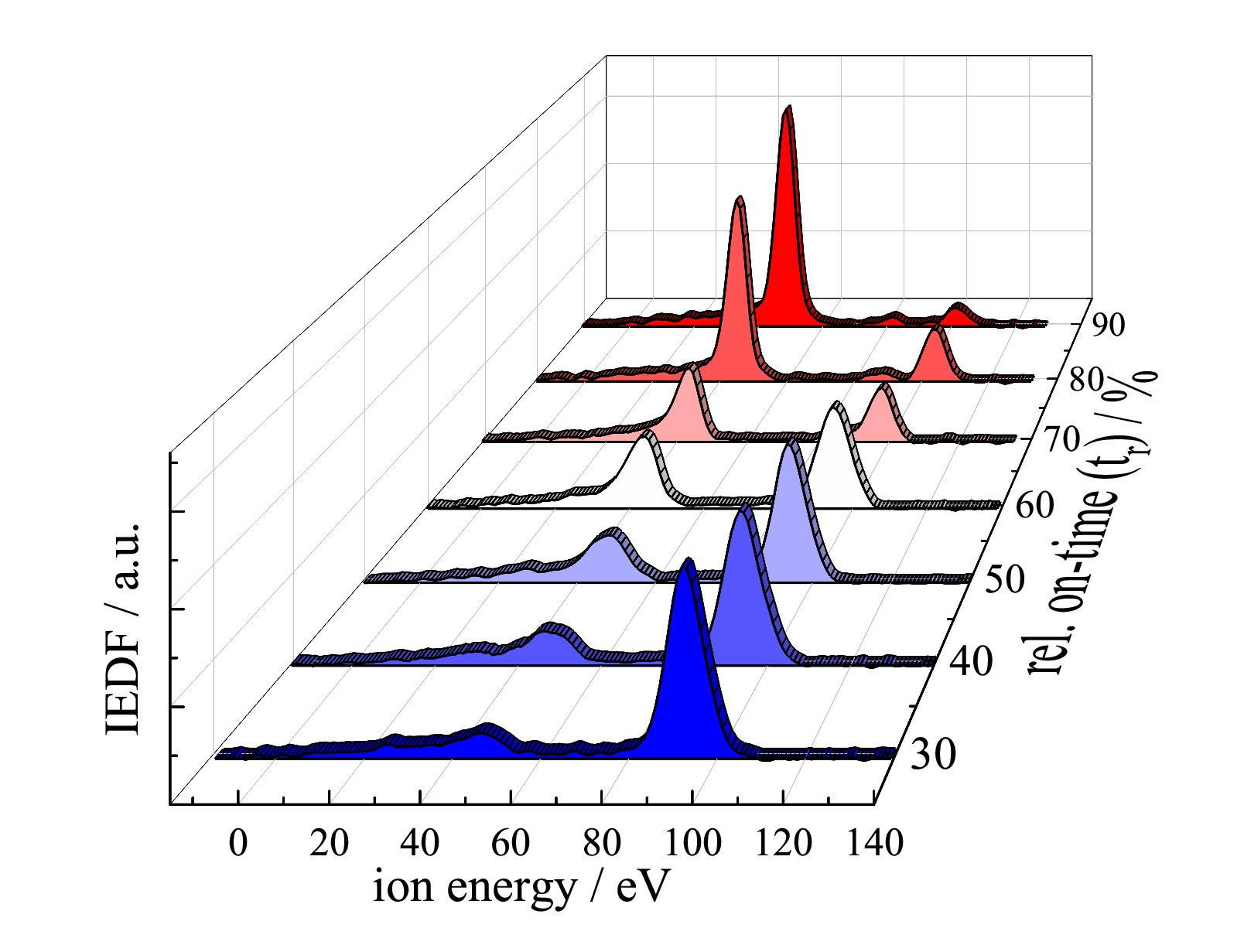}
            \caption{}
            \label{fig:relT}   
            
        \end{subfigure}
        \caption{IEDFs at the substrate electrode measured in argon at \SI{100}{\watt} ICP source power at \SI{1}{\pascal}. The triple-plateau waveform was used with (a) a varying relative voltage, $V_r$ at $t_r$ = \SI{75}{\percent}, and (b) a varying relative on-time, $t_r$ at $V_r$ = 0.4. }
        \label{fig:dualPeakAr}
\end{figure}

In figure \ref{fig:dualPeakAr}, the measured IEDFs at the substrate electrode (Ar, 1 Pa, 100 W ICP source power) are shown in the presence of the triple-plateau VWT substrate bias as a function of $V_r$ and $t_r$.

In figure \ref{fig:relV} the substrate voltage decreased from \SI{122}{\volt} peak-to-peak
at Vr = 0.7 to \SI{94}{\volt} at Vr = 0.1. The changing output voltage of the amplifier indicates a change in the load resistance of the plasma. In figure \ref{fig:relT} the substrate voltage stayed constant at \SI{100}{\volt} peak-to-peak. 
Figure \ref{fig:relV} confirms that $V_r$ mainly affects the energy gap between the two IEDF peaks. However, the flux ratio also changes with $V_r$. At $V_r = 0.7$, the low energy peak is much stronger than the high energy peak, while at $V_r = 0.1$, both peaks are approximately equal. In figure \ref{fig:relT} the effects of varying $t_r$ are shown. The data show that the flux ratio between the two peaks can be precisely adjusted using this control parameter. The position of the IEDF peaks on the energy axis is not affected by the variation of $t_r$. These results suggest that any desired IEDF shape can be generated by splitting it into individual peaks and adding additional plateaus to the voltage waveform.
To realize such a controllable dual-peak IEDF a total of six voltage waveform control parameters must be adjusted (duty cycle, absolute peak-to-peak value, relative voltage, relative on-time, slope of the first plateau, slope of the second plateau). Each additional peak would add another three waveform control parameters for the additional waveform plateau (slope, relative voltage, and relative on-time). 
The dual plateau waveform was also tested in SF\textsubscript{6}, showing comparable results. Exemplary measured IEDFs can be found in the supplementary material.

\subsection{Spatio-temporal electron dynamics and density\\}
After focussing on the ion dynamics, in this section the influence of the LF VWT bias on the electron dynamics and density is investigated. 

\begin{figure}[!h]%here, top, bottom
        \centering
        \includegraphics[width=0.8\textwidth,trim=40 00 40 0, clip]{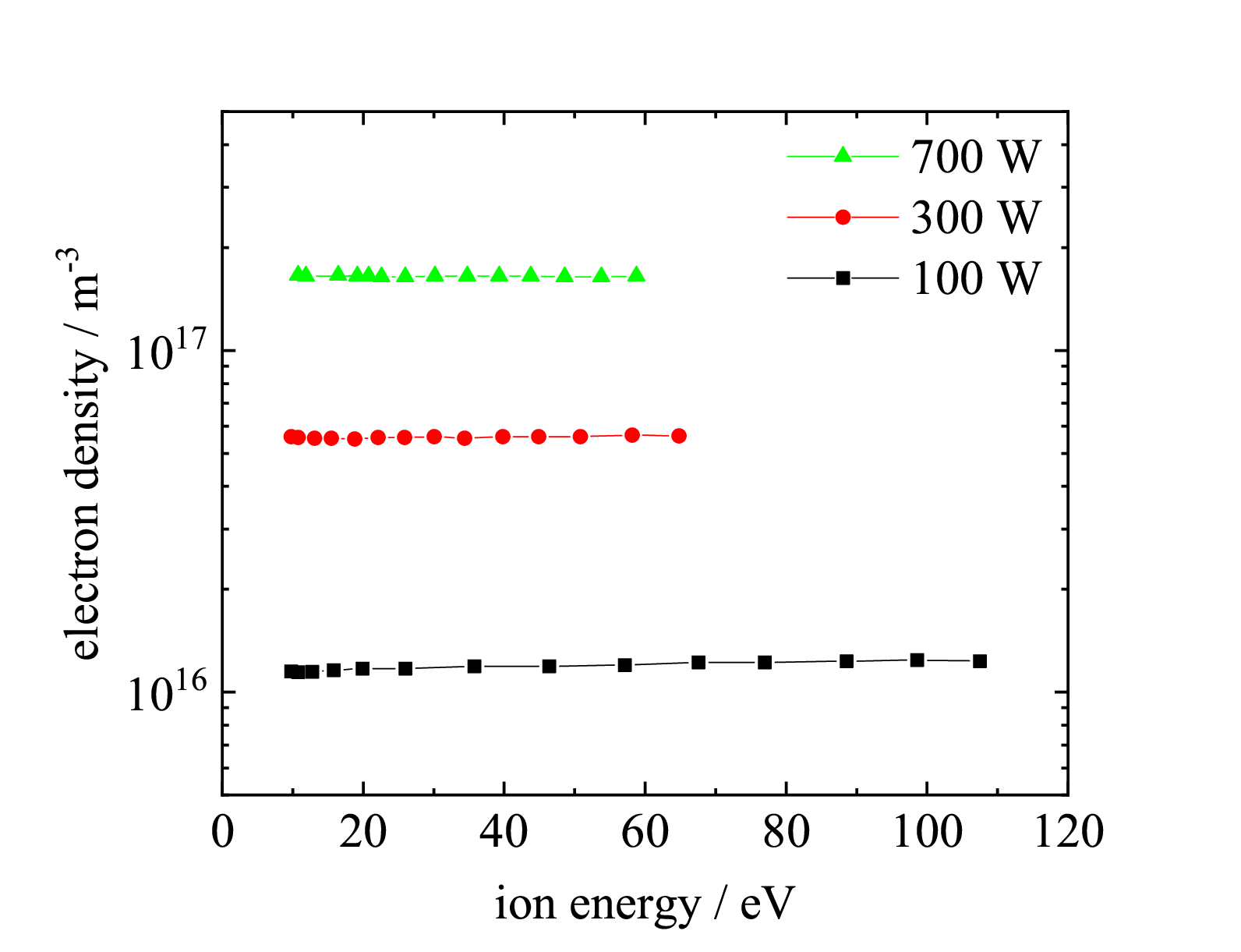}
        \caption{Electron density as a function of the target ion energy using the LF VWT bias in argon at \SI{1}{\pascal} and \SIlist{100;300;700}{\watt} ICP source power. The electron density is measured by a MRP in the center of the discharge, \SI{1}{\centi\meter} above the substrate.}
        \label{fig:MRP}
\end{figure}

Figure \ref{fig:MRP} shows the electron density as a function of the peak ion energy controlled via the LF VWT substrate bias peak-to-peak voltage for \SI{100}{\watt}, \SI{300}{\watt} and \SI{700}{\watt} ICP source power. The electron density remains nearly constant throughout the entire parameter range. Only at \SI{100}{\watt}, the electron density increases by \SI{7}{\percent} from \SI[per-mode = reciprocal]{1.15e16}{\per\cubic\meter} to \SI[per-mode = reciprocal]{1.23e16}{\per\cubic\meter} as a function of the ion energy. This indicates a weak effect of the subtrate bias on the plasma density at low ICP source power.

\label{sec:electronDynamics}
\begin{figure*}[!h]%here, top, bottom
        \centering
        \begin{subfigure}{0.45\textwidth}
        \includegraphics[width=\textwidth,trim=0 00 0 0, clip]{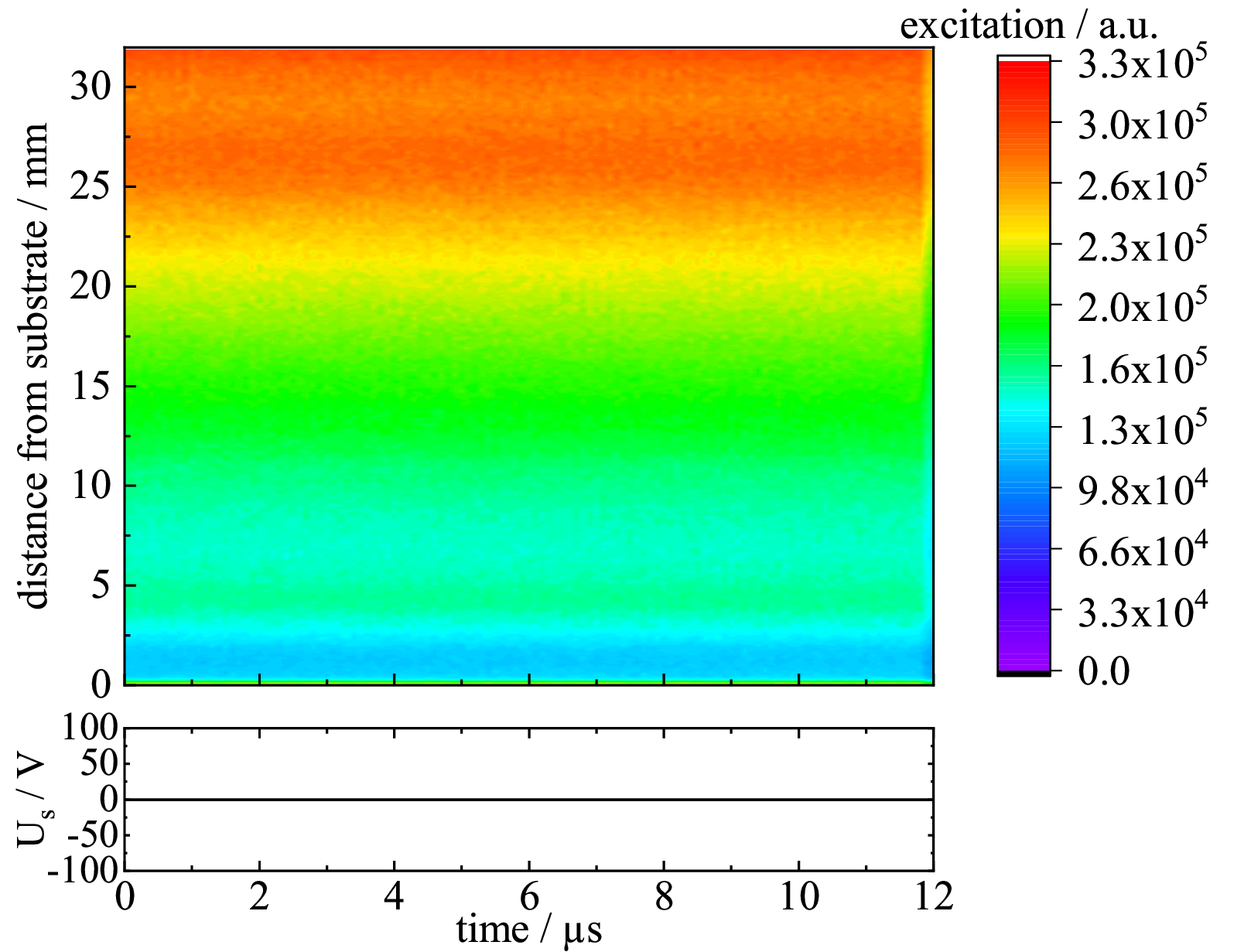}
        \caption{\SI{0}{\volt} LF VWT substrate bias}
        \label{fig:p0mV}
        \end{subfigure}
        %\hspace*{\fill}
        \begin{subfigure}{0.45\textwidth}
        \includegraphics[width=\textwidth,trim=0 00 0 0, clip]{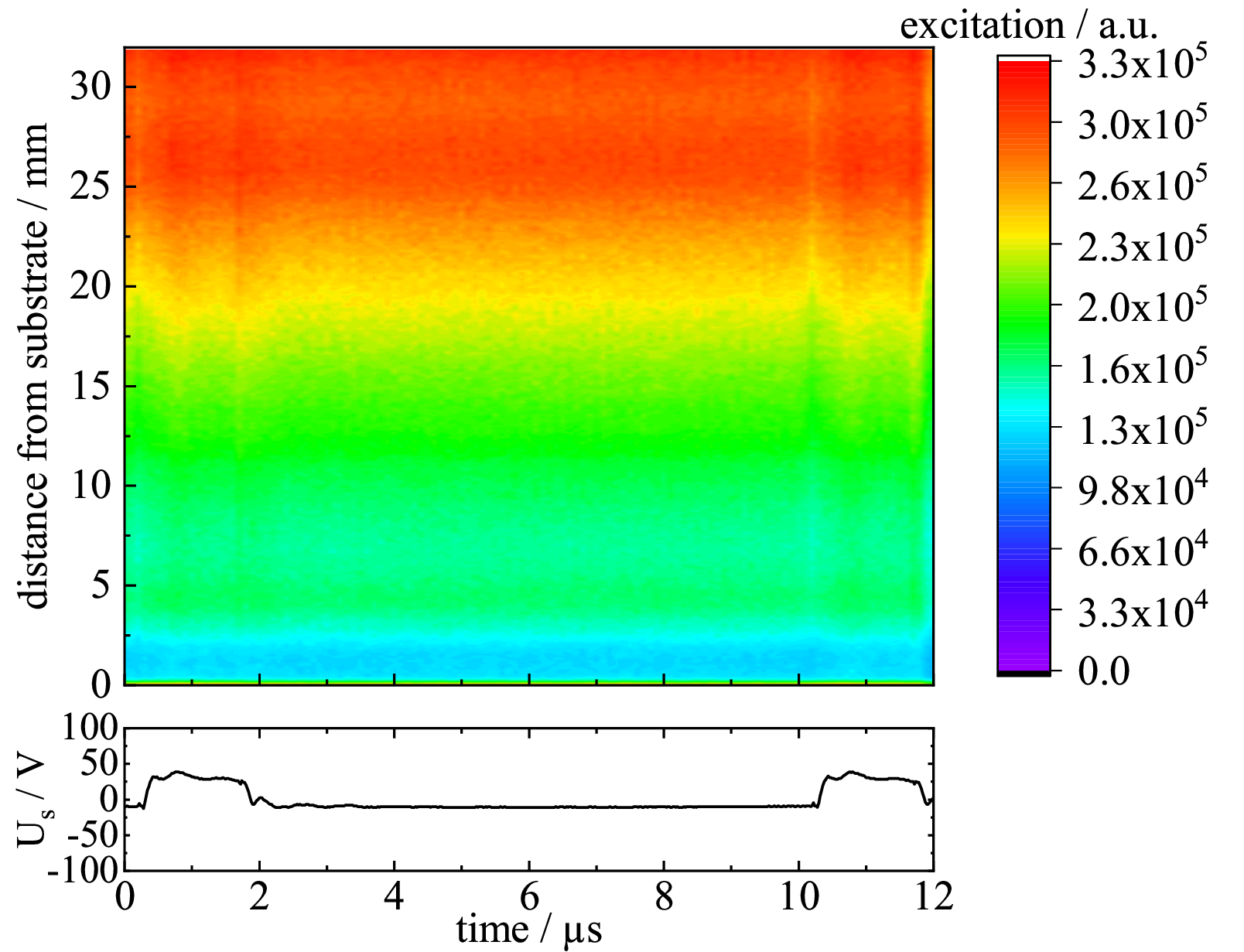}
            \caption{\SI{50}{\volt} LF VWT substrate bias}
            \label{fig:p40mV}   
        \end{subfigure}
        %\hspace*{\fill}
        \begin{subfigure}{0.45\textwidth}
        \includegraphics[width=\textwidth,trim=0 00 0 0, clip]{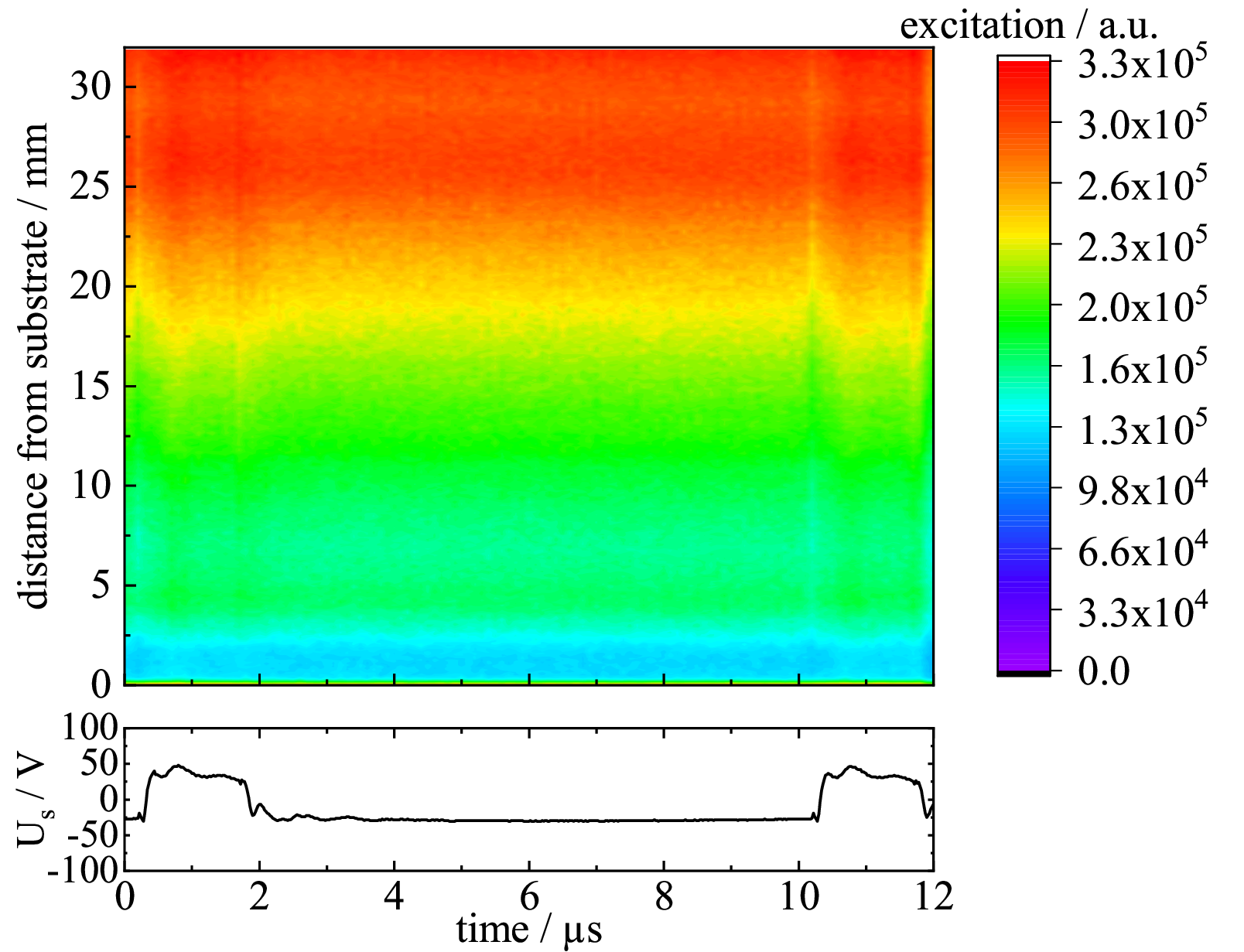}
            \caption{\SI{80}{\volt} LF VWT substrate bias}
            \label{fig:p60mV}   
        \end{subfigure}
        %\hspace*{\fill}
        \begin{subfigure}{0.45\textwidth}
        \includegraphics[width=\textwidth,trim=0 00 0 0, clip]{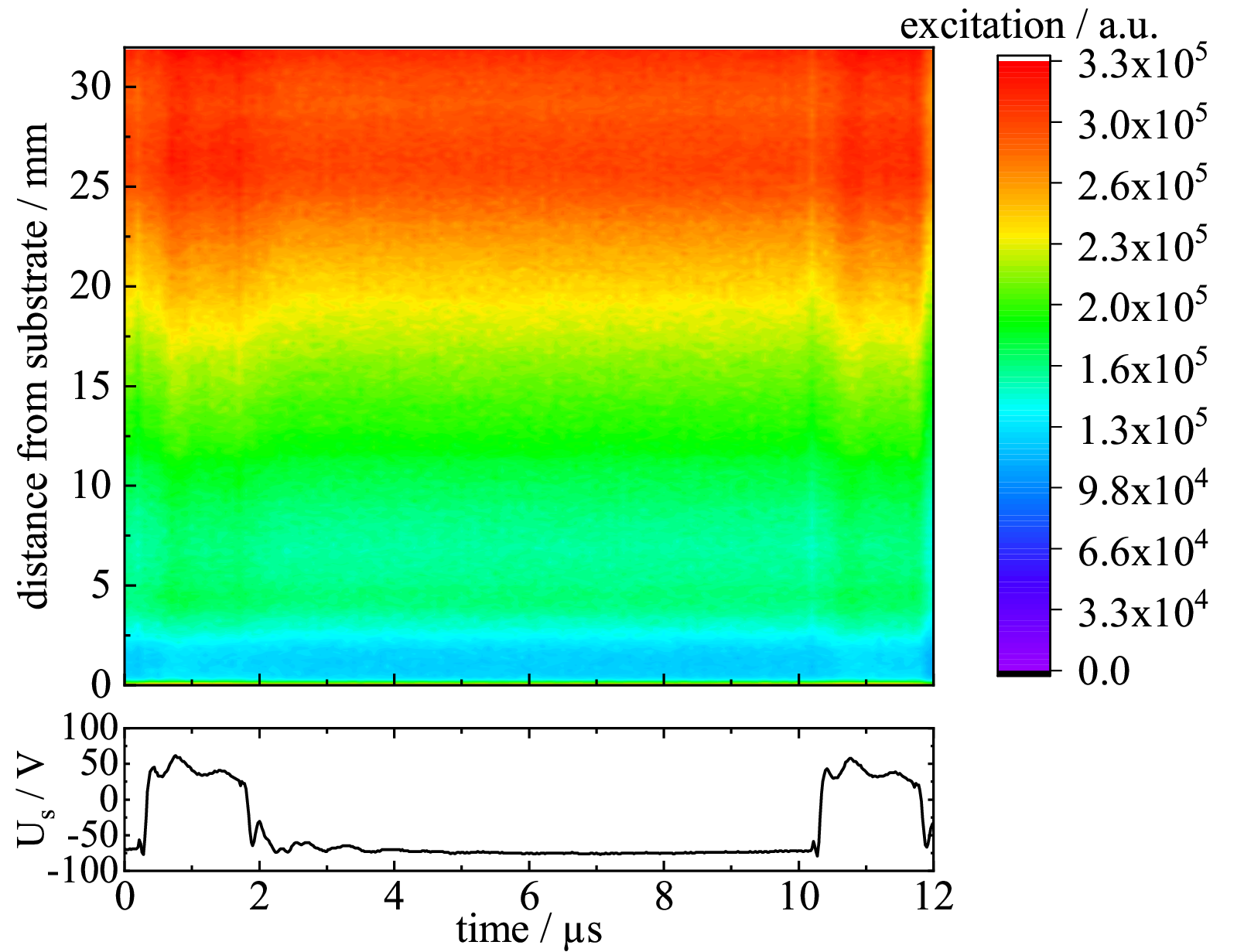}
            \caption{\SI{140}{\volt} LF VWT substrate bias}
            \label{fig:p100mV}   
        \end{subfigure}
        \caption{Spatio-temporally resolved electron impact excitation rate from the ground state into the Ar 2p\textsubscript{1} level in Ar at \SI{100}{\watt} ICP source power, \SI{1}{\pascal} for different LF substrate peak-to-peak voltages. For each peak-to-peak voltage, the corresponding substrate voltage waveform is also shown. }
        \label{fig:PROES}
\end{figure*}

Figure \ref{fig:PROES} shows the spatio-temporally resolved electron impact excitation rate from the ground state into the Ar 2p\textsubscript{1} level at \SI{100}{\watt} ICP source power at \SI{1}{\pascal} for different LF VWT substrate peak-to-peak voltages in a spatial region close to the substrate. Below each panel, the corresponding substrate voltage waveform, $U_s$, is shown. The horizontal axis ranges over 1.2 periods of the bias waveform. \\
In figure \ref{fig:p0mV} the background excitation caused by acceleration of electrons by the azimuthal electric field induced by the inductive coupling of the coil is shown. As the camera is only triggered by the LF VWT substrate bias waveform, no temporal resolution of the electron dynamics caused by the inductive coupling is obtained. In figure \ref{fig:PROES} the substrate surface itself is not visible, as the line of sight of the camera is blocked by the wafer clamping ring, which is \SI{2}{\milli\meter} thick.
The excitation rate drops as the distance from the ICP coil increases (i.e., the distance to the substrate decreases), since the field of view is below the skin layer of the ICP \cite{Mahoney1994}. 
At the heights of \SI{7}{\milli\meter} and \SI{29}{\milli\meter} a small drop in the reported excitation rate is visible, which is caused by optical reflections in the diagnostic system. Figures \ref{fig:p40mV}–\ref{fig:p100mV} show the excitation rate at different LF VWT substrate peak-to-peak voltages. During the positive part of the waveform (\SIrange[]{1,4}{1,9}{\micro\second}), two emission peaks can be observed. On the rising edge, the sheath collapses, and electrons are accelerated towards the substrate to compensate the positive ion flux to the substrate on time average \cite{Schulze_2008}.

During the falling edge of the substrate voltage, the sheath expands and electrons are accelerated into the bulk, creating the second peak. This emission peak decays rapidly as the electron density in the expanding sheath drops. During the negative part of the substrate voltage (\SIrange[]{2}{10}{\micro\second}), the ion flux towards the substrate generates secondary electrons. These $\gamma$-electrons are accelerated by the sheath electric field towards the plasma bulk and lead to higher excitation than in the case of 0 V substrate bias.

The peak ion energy at the substrate changes from \SI{25}{\electronvolt} to \SI{85}{\electronvolt} by changing the substrate peak-to-peak voltage from 0 V to 140 V. For all substrate peak-to-peak voltages, the spatio-temporally resolved electron impact excitation dynamics is similar indicating that the LF VWT bias hardly affects the dynamics of energetic electrons and, thus, the electron density. An increase of the ICP source power does not change the qualitative behavior of the electron dynamics. Data for \SIlist{100;200;300;400}{\watt} source power with \SI{60}{\electronvolt} bias can be found in the supplementary material.

\subsection{Sputter etch rates}
\label{sec:etchting}
\begin{figure}[h]%here, top, bottom
        \centering
        \includegraphics[width=0.8\textwidth,trim=40 00 40 0, clip]{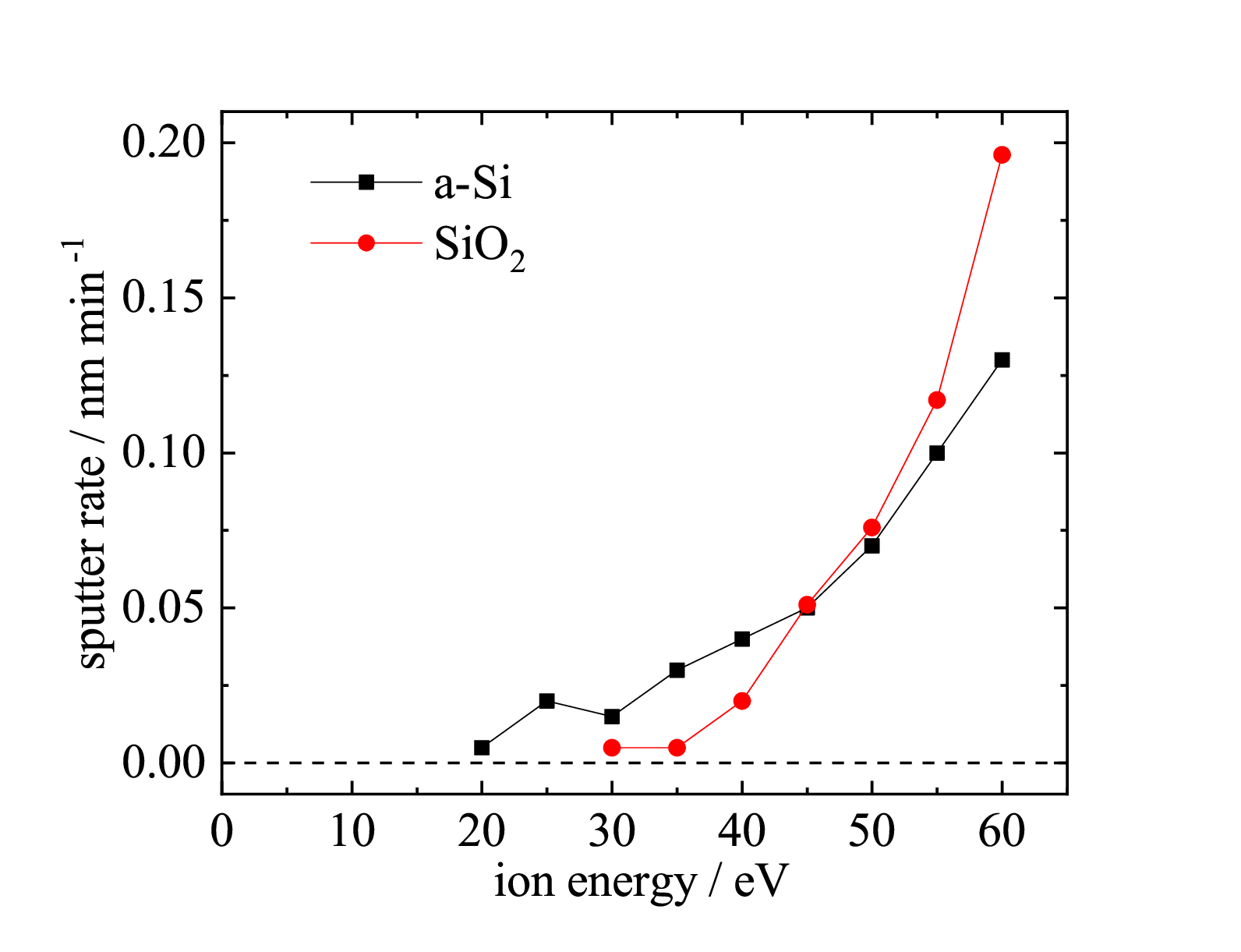}
        \caption{Sputter etch rates for a-Si and \sd{} as a function of the peak ion energy, controlled by varying LF VWT substrate peak-to-peak voltage, in argon at \SI{1}{\pascal}, \SI{100}{\watt} ICP source power.}
        \label{fig:RatesVWT}
\end{figure}

Sputter etch rates have been measured for silicon dioxide (\sd{}) and amorphous silicon. The wafers were exposed to an argon plasma at \SI{1}{\pascal}, \SI{100}{\watt} ICP source power and at different peak ion energies for 10 minutes, respectively. The peak ion energy was set using the closed-loop VWT bias control based on equation \ref{eq:Ei}. The system reached steady state with no overshooting after a few seconds. Thus, the settling time can be ignored as it is much shorter than the plasma exposure time. 
To test the validity of the energy prediction on a wafer, a single RFEA sensor was placed on a dummy wafer and the peak ion energy was measured under the same process conditions. The measured peak ion energies were consistently between \SI{1}{\electronvolt} and \SI{2}{\electronvolt} lower than the predicted energy according to equation \ref{eq:Ei} due to the voltage drop across the oxide layer on the wafer. Because this energy shift is within the margin of error, it is not taken into account here.
The wafer with the a-Si layer was treated with 20 minutes of plasma exposure with \SI{120}{\electronvolt} peak ion energy to remove the native oxide layer before the first measurement. 
An Al\textsubscript{2}O\textsubscript{3} wafer was also placed onto the substrate electrode, but was not etched by ion bombardment at energies accessible in this setup. This suggests that the sputter threshold of Al\textsubscript{2}O\textsubscript{3} exceeds the maximum ion energy accessible in our setup of \SI{130}{\electronvolt}. Day \etal \cite{Day_1992} reported sputtering of Al\textsubscript{2}O\textsubscript{3} at ion energies of \SI{100}{\electronvolt}, however, they employed a sinusoidal bias voltage at \SI{13.56}{\mega\hertz} and used the sum of the DC bias voltage and the plasma potential as the reported ion energy. As seen above, the actual energy of the fastest ions is typically much larger. The publication does not give any further information about the IEDF shape. Taking this into account, the presented findings are in agreement with previous work.
Figure \ref{fig:RatesVWT} shows the obtained etch rates for \sd{} and silicon. 
The measurements for \sd{} show no sputtering for energies up to \SI{35}{\electronvolt}. For higher energies, a rising sputter rate in the range of \SI{0.025}{\nano\meter\per\minute} to \SI{0.2}{\nano\meter\per\minute} is found. A linear fit in the range of \SI{40}{\electronvolt} to \SI{55}{\electronvolt} indicates a sputter threshold of \SI{37}{\electronvolt}. This is in good agreement with the works of Faraz \etal \cite{faraz2020precise} and Wei \etal \cite{Wei_2025} who reported \SI{36}{\electronvolt} and \SI{37}{\electronvolt}, respectively. Both works use mono-energetic IEDFs. For a-Si significant sputtering already occurs at lower energies. Here, a linear fit in the range of \SI{30}{\electronvolt} to \SI{45}{\electronvolt} yields a threshold ion energy of \SI{23}{\electronvolt}. Little data is available for sputtering yields of silicon at low ion energies. Wolsky and Zdanuk \cite{Wolsky_1961} reported sputtering yields for mono-crystalline silicon in the range of \SI{34}{\electronvolt} to \SI{800}{\electronvolt}. By extrapolating the low-energy data points, they predict a sputtering threshold between \SI{15}{\electronvolt} and \SI{20}{\electronvolt}. Barone and Graves \cite{Barone_1995} used molecular dynamics simulations to investigate the physical and chemical sputtering of fluorinated silicon. They report a threshold of \SI{20}{\electronvolt} and \SI{4}{\electronvolt} for physical sputtering and chemical sputtering, respectively. However, this was done for a highly fluorinated surface and with only 3 data points. The authors also reported a strong influence of the F/Si ratio on both sputtering yields at higher ion energies. So, despite the separation of physical and chemical sputtering, the results cannot be directly compared. Shibanov \etal \cite{Shibanov_2023} also measured the sputter yield of amorphous silicon in Argon, Xenon and Krypton plasma. Despite them using a \SI{12}{\mega\hertz} sinusoidal bias voltage, they found process conditions where the superposition of the applied voltage and the harmonics generated by the plasma led to an essentially mono-energetic IEDF. In their work, they found a non-zero sputter yield for a-Si bombarded by Ar ions at \SI{22}{\electronvolt}, which was the lowest reported ion energy. 

Based on these results, generating narrow and highly controllable IEDFs via LF VWT substrate biasing, allows realizing a high degree of selectivity in etching these two materials in pure Argon plasmas via sputtering. If the IEDF is narrow and tuned to have its peak above the sputter threshold of a-Si and below that of \sd{}, a high selectivity will be realized. This approach can be extended to other material combinations, enabling sustainable and selective etching processes by using only inert gases. The importance of VWT becomes even more important for higher threshold energies, because of the increasing width of the conventionally generated IEDFs at high substrate voltages.

\section{Conclusions}

In this work, the influence of a low-frequency tailored waveform substrate bias on a variety of plasma and process parameters was investigated in a commercial RIE reactor. The goal of controlling the IEDF shape by LF VWT substrate biasing without affecting other parameters such as electron density and ion flux was successfully achieved. With a pulse-wave-shaped voltage waveform mono-energetic IEDFs with a FWHM below \SI{10}{\electronvolt} and a tunable ion energy were generated. It was found that the ion energy could be predicted non-intrusively by measuring the time-resolved bias voltage. Using this approach, a simple control algorithm was developed to compensate for drifting process parameters. 
At a low ICP source power, the ion flux was unaffected by the bias voltage, whereas at higher ICP source powers the ion flux increases with increasing bias voltages up to an ICP source power-dependent saturation point at which the ion flux stays constant. The peak ion energy at which the ion flux saturates increases with increasing ICP source power. 
By adding another voltage plateau to the LF VWT bias waveform, two independently controllable peaks of the IEDF could be generated. Both their position on the energy axis and their height can be controlled by the shape of the substrate bias voltage waveform. The influence of the tailored bias on the electron dynamics and density was found to be low, as indicated by PROES and MRP measurements. The sputter threshold energies for argon ions were determined to be \SI{37}{\electronvolt} for \sd{} and \SI{23}{\electronvolt} for a-Si, thereby confirming the presence of a viable selectivity window. Generating narrow IEDFs via VWT, thus, allows realizing selective plasma etching of these and potentially other materials. 
These results demonstrate the advantages of LF VWT substrate biasing for selective plasma etching via IEDF shape control. These insights provide the basis for future applications of VWT for selective and efficient etching of other materials, including Atomic Layer Etching (ALE).

\section*{Acknowledgements}
This work was supported by the Federal Ministry of Research, Technology and Space of Germany (BMFTR) within the ForMikro project FlexTMDSense (grant number 16ES1096K) and the ForLab PICT2DES (grant number 16ES0941). The authors thank the technical staff of the Microsystems Technology chair (MST) for the support in the cleanroom and the team of the Center for Interface-Dominated High Performance Materials (ZGH) for carrying out the deposition process of a-Si. 

\newpage

\bibliographystyle{iopart-num}
\bibliography{bibliography}

\end{document}